\documentclass[aps]{revtex4}
\usepackage{graphicx}

\begin{document}

\title{Revisiting the anomalous rf field penetration into a warm plasma}
\author{Igor D. Kaganovich }
\affiliation{Plasma Physics Laboratory, Princeton University, Princeton, NJ 08543}
\author{Oleg V. Polomarov, Constantine E. Theodosiou}
\affiliation{Department of Physics and Astronomy, University of Toledo, Toledo, Ohio,
43606-3390.}
\date{\today }

\begin{abstract}
Radio frequency waves do not penetrate into a plasma and are damped within
it. The electric field of the wave and plasma current are concentrated near
the plasma boundary in a skin layer. Electrons can transport the plasma
current away from the skin layer due to their thermal motion. As a result,
the width of the skin layer increases when electron temperature effects are
taken into account. This phenomenon is called anomalous skin effect. The
anomalous penetration of the rf electric field occurs not only for
transversely propagating to the plasma boundary wave (inductively coupled
plasmas) but also for the wave propagating along the plasma boundary
(capacitively coupled plasmas). Such anomalous penetration of the rf field
modifies the structure of the capacitive sheath. Recent advances in the
nonlinear, nonlocal theory of the capacitive sheath are reported. It is
shown that separating the electric field profile into exponential and
non-exponential parts yields an efficient qualitative and quantitative
description of the anomalous skin effect in both inductively and
capacitively coupled plasma.
\end{abstract}

\maketitle

\section{Introduction}

A radio frequency electromagnetic wave does not penetrate into a plasma if
the wave frequency $\omega $ is smaller than the electron plasma frequency $%
\omega _{p}=\sqrt{4\pi e^{2}n_{e}/m}$, where $e$ and $m$ are the electron
charge and mass, respectively, and $n_{e}$ is the electron density.
Electrons distribute their charge and current so as to shield out the
electromagnetic wave. The shielding depends on the direction of the wave
with regard to the plasma boundary. If the wave electric field is
perpendicular to the plasma boundary, the rf field penetrates into the
plasma only within a depth of the order of the Debye length $v_{T}/\omega
_{p}$, where $v_{T}=\sqrt{2T_{e}/m}$ is the electron thermal velocity,
determined by the electron temperature $T_{e}$, in eV. If the wave electric
field is along the plasma boundary, the rf field penetrates into the plasma
only within a depth of the order of the skin depth $c/\omega _{p}$, where $c$
is the speed of light in vacuum. Here, we consider a ``collisionless"
plasma, i.e. where the collision frequency is small compared to the wave
frequency $\nu \ll \omega$ and the electrons undergo rare collisions during
the rf cycle; thus, collisions have little effect on wave screening by
plasma.

\begin{figure}[th]
\begin{center}
\includegraphics[width=3in]{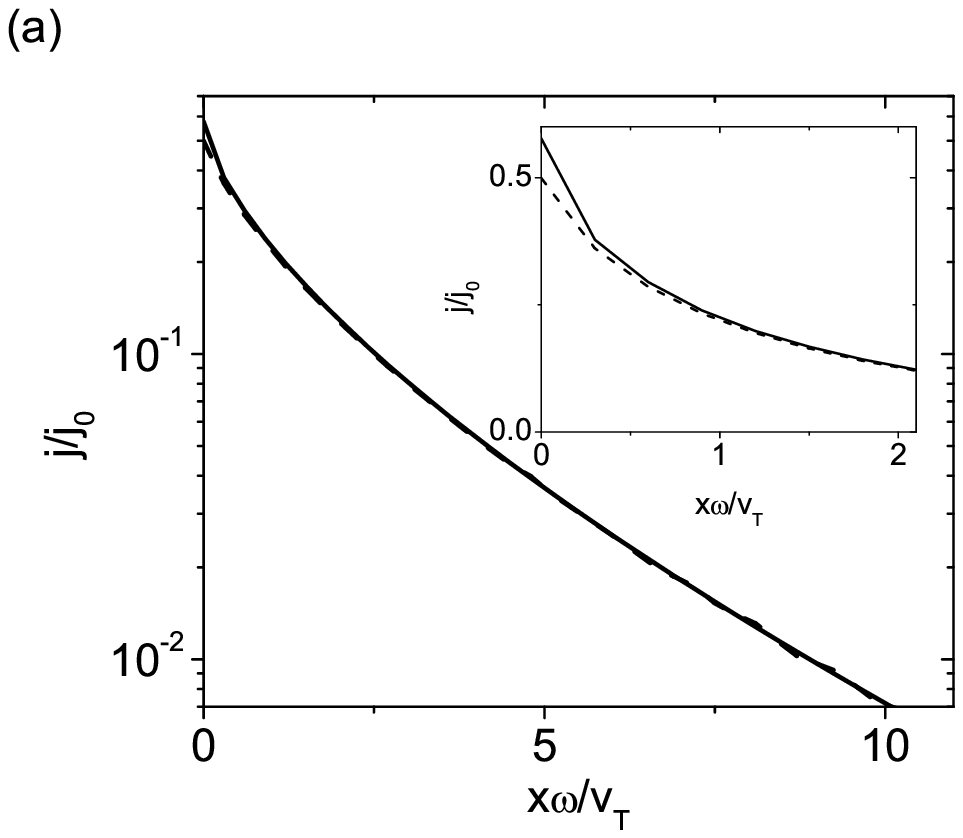} %
\includegraphics[width=3in]{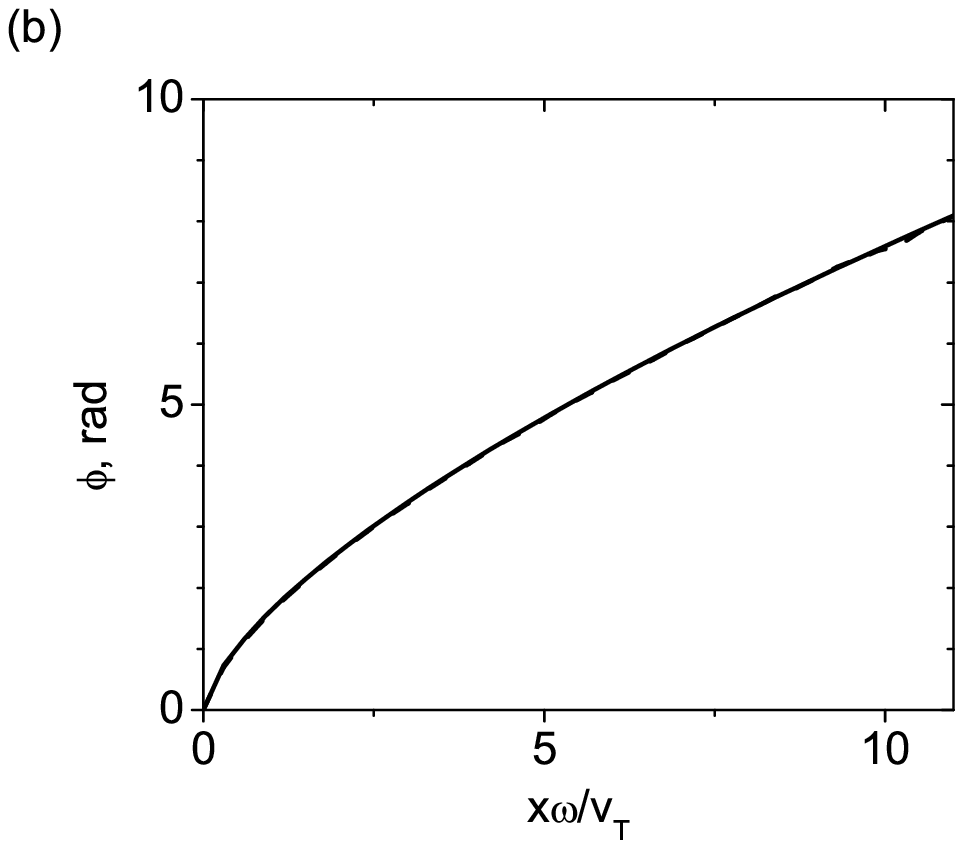}
\end{center}
\caption{{}Phase-mixing of the test particle current generated by velocity
kicks $\Delta V\cos (\protect\omega t)$ at the plasma boundary: (a) current
amplitude and (b) the current phase with respect to the phase of the
velocity kick at the plasma boundary. The amplitude of the current is
normalized on $j_{0}=en_{0}\Delta V$, where $n_{0}$ is plasma density. Solid
lines show the exact result of numerical integration in Eq.~(\protect\ref%
{current integral}), dashed lines show the asymptotic, approximate
analytical results given by Eq.~(\protect\ref{current approx}).}
\label{Fig decay}
\end{figure}

Another important scale is the nonlocality or phase-mixing scale $%
v_{T}/\omega $, which determines the scale length of the electron current
profile in the plasma. To demonstrate the concept of phase-mixing scale $%
v_{T}/\omega $ let us consider a simple model, where an electron acquires a
prescribed velocity kick near the plasma boundary, in the direction
perpendicular to the boundary%
\begin{equation}
dv_{x}(t)=\Delta V\exp (-i\omega t).
\end{equation}%
The electron velocity at a distance $x$ from the boundary will be determined
by the moment when velocity kick was acquired at the plasma boundary, i.e.,
by the time $t-x/v_{x}$. The electron current in the plasma is given by an
integration over all electrons with a velocity distribution function $%
f(v_{x})$
\begin{equation}
j(x,t)=e\Delta V\int_{0}^{\infty }f(v_{x})\exp [-i\omega (t-x/v_{x})]dv_{x}.
\label{current}
\end{equation}%
Here, only electrons collided with the wall ($v_{x}>0$) have to be taken
into account. For a Maxwellian distribution function $%
f(v_{x})=n_{0}e^{-v_{x}^{2}/v_{T}^{2}}/v_{T}\sqrt{\pi }$, the plasma current
in Eq.(\ref{current}) becomes
\begin{equation}
j(x,t)=\frac{j_{0}e^{-i\omega t}}{\sqrt{\pi }}\int_{0}^{\infty }\exp \left(
-s^{2}+\frac{i\omega x}{v_{T}s}\right) ds,  \label{current integral}
\end{equation}%
where $s=v_{x}/v_{T}$ and $j_{0}=en_{0}\Delta V$. The amplitude and phase of
the current are shown in Fig.1. In the limit $\omega x/v_{T}\gg 1$, the
integration in Eq.~(\ref{current integral}) can be performed analytically
making use of the method of steepest descend \cite{Brillouin}, see Appendix
A for more details. This gives
\begin{equation}
j(x,t)\approx \frac{j_{0}}{\sqrt{3}}\exp \left[ -i\omega t-\frac{3}{4}\left(
\frac{x}{\lambda _{\omega }}\right) ^{2/3}+i\frac{3\sqrt{3}}{4}\left( \frac{x%
}{\lambda _{\omega }}\right) ^{2/3}\right] ,  \label{current approx}
\end{equation}%
where $\lambda _{\omega }=v_{T}/\sqrt{2}\omega $ is the phase-mixing scale.
Comparison of the asymptotic calculation result given by Eq.~(\ref{current
approx}) with the exact result of numerical integration in Eq.~(\ref{current
integral}) is shown in Fig.~\ref{Fig decay}. From Fig. \ref{Fig decay}, it
is evident that Eq.~(\ref{current approx}) approximates the exact result for
any $x$ within a 15 percent error bar. The largest error occurs at $x=0$,
where half of the electron population with velocity $v_{x}>0$ acquired the
velocity kick, which gives rise to the electron current $j(0)=j_{0}/2$,
whereas Eq.~(\ref{current approx}) predicts $j(0)=j_{0}/\sqrt{3}$, which
corresponds to a 15 percent error.

Equation (\ref{current approx}) describes the process of phase mixing -
electrons with velocities different by $\delta v_{x}\sim v_{T}$ have
different phase lag of the order $\omega x/v_{T}$ at a distance $x$ from the
plasma boundary. Therefore, at $x\sim v_{T}/\omega $ the phase difference
becomes considerable: contributions to the total current from electrons with
different velocities $v_{x}$ cancel out each other, and the plasma current
vanishes. Interestingly, the spatial profile of the current is not a simple
exponential function, but an exponential function of $\left( x/\lambda
_{\omega }\right) ^{2/3}$. As it will be shown below this is typical for the
spatial profiles of the electric field and electron current in warm plasmas
due to nonlocal effects.

So far, we solved only test-particle problem and did not take into account
the plasma polarization. The current in Eq.~(\ref{current integral}) is
nonuniform; thus, there must be an electron density perturbation according
to the continuity equation
\begin{equation}
e\frac{\partial n_{e}}{\partial t}=-\frac{\partial j}{\partial x}.
\end{equation}%
The electron density perturbations polarize the plasma and generate an
electric field, which in turn, affects the electron motion and the electron
current profile. Thus, Eq.(\ref{current integral}) has to be modified to
include the self-consistent electric field. This requires solving the Vlasov
equation together with the Poisson equation. In his famous 1946 paper,
Landau obtained an analytic solution for the penetration of the longitudinal
rf electric field into a plasma \cite{Landau}. Note that he also described
\textquotedblleft Landau damping" in the same paper. We briefly review his
solution for a small amplitude electric field in the linear approximation
and discuss the more realistic case of a large amplitude electric field.

The structure of this review is as follows: In section II, the penetration
of the longitudinal electric field into the plasma is described. This case
corresponds to a capacitively coupled plasma. In section III, the
penetration of the transverse electric field into the plasma is studied,
which corresponds to an inductively coupled plasma. In subsection III.E, it
is shown that anisotropy of the electron velocity distribution function can
have a profound effect on the anomalous skin effect.

\section{Penetration of the rf electric field directed perpendicular to the
plasma boundary (capacitively-coupled plasma)}

\subsection{Small-amplitude electric field}

In the previous section, we considered a test particle current driven by
artificially applied velocity modulations at the plasma boundary. Here,
self-consistent penetration of a small amplitude rf electric field directed
perpendicular to the plasma boundary is considered. Such a model provides
some insight into the sheath structure of capacitively-coupled plasmas.

First, let's consider a stationary negatively biased electrode. It is
well-known that the externally applied electric field penetrates inside the
plasma over distances of the order of the Debye length $a=v_{T}/\sqrt{2}%
\omega _{p}=\sqrt{T_{e}/4\pi e^{2}n_{0}}$. The plasma electrons are trapped
by the plasma potential, $\phi (x)$, in the potential well $-e\phi (x)$. The
electron density obeys the Boltzmann distribution
\begin{equation}
n_{e}(x)=n_{0}\exp \left[ e\phi (x)/T_{e}\right].  \label{Boltzmann}
\end{equation}%
The Poisson equation
\begin{equation}
\frac{d^{2}\phi }{dx^{2}}=-4\pi e(n_{i}-n_{e})  \label{Poisson general}
\end{equation}%
can be simplified assuming small potential variations $-e\phi (x)/T_{e}\ll 1$
and a uniform background plasma with $n_{e}=n_{i}=n_{0}$. Thus, Eq.~(\ref%
{Poisson general}) becomes
\begin{equation}
\frac{d^{2}\phi }{dx^{2}}=\frac{4\pi e^{2}n_{0}}{T_{e}}\phi .
\label{Poisson linear}
\end{equation}%
The solution of Eq.~(\ref{Poisson linear}) is an exponentially decaying
electric field $E=-d\phi /dx$
\begin{equation}
E=E_{0}\exp \left( -\frac{x}{a}\right) .
\end{equation}%
Here, $E_{0}$ is the value of the electric field at the plasma boundary.
This is the solution for a steady state, time-independent sheath electric
field. In the opposite case of the time-dependent electric field, the
Boltzmann distribution given by Eq.~(\ref{Boltzmann}) is no longer valid and
the electron density has to be determined from the Vlasov equation. Landau
solved the Vlasov equation coupled with the Poisson equation analytically in
the linear approximation considering an electrostatic wave with small
amplitude $|e\phi (x)|/T_{e}\ll 1$ and small frequency $\omega \ll \omega
_{p}$ \cite{Landau}. Details of the solution are described in Appendix B.

\begin{figure}[th]
\begin{center}
\includegraphics[width=3in]{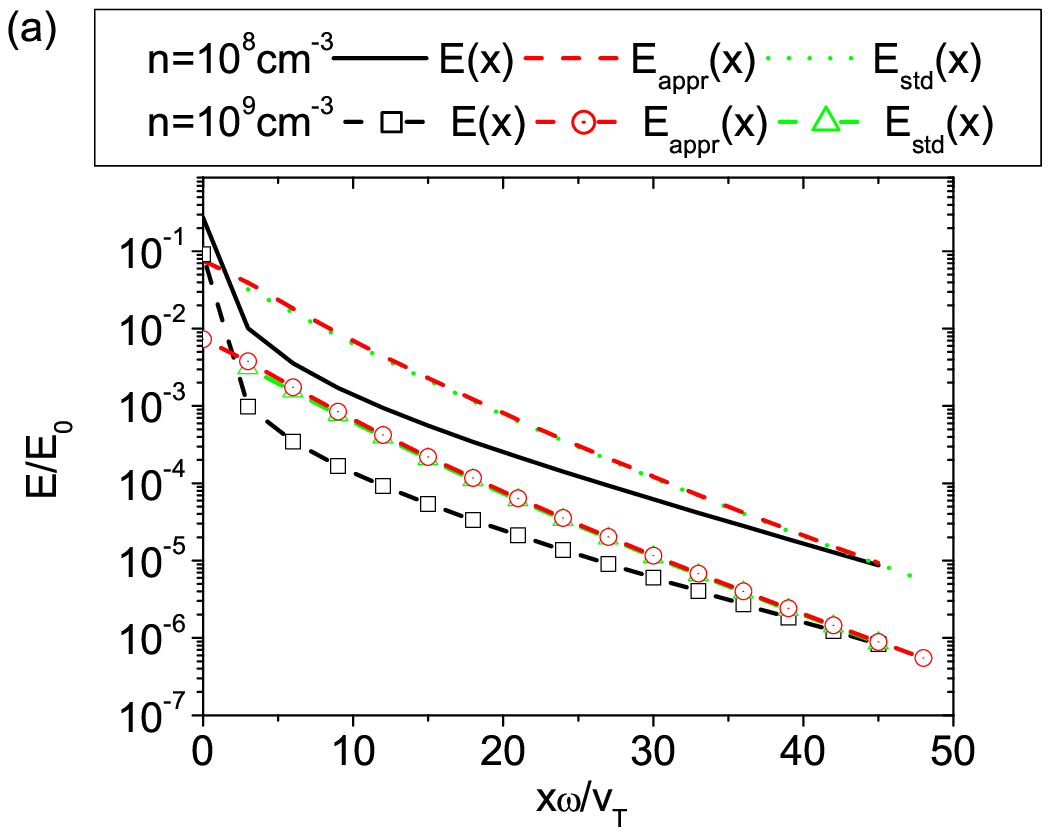} %
\includegraphics[width=3in]{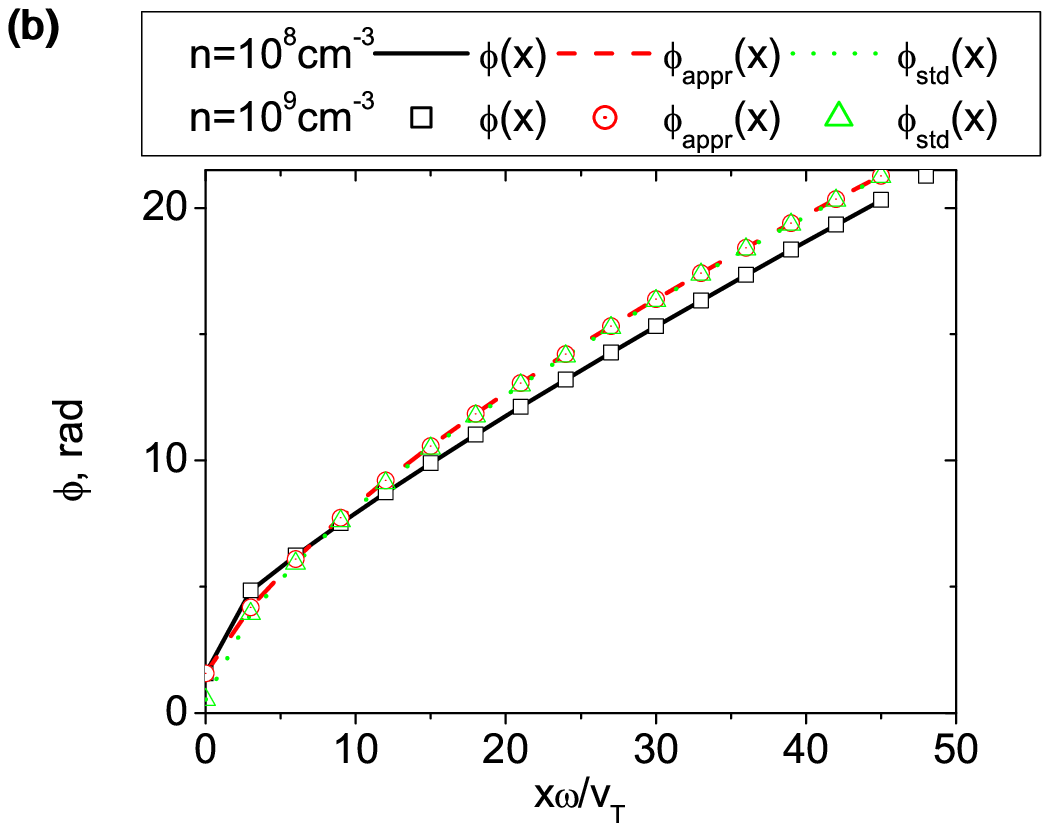}
\end{center}
\caption{{} Penetration of the external electric field into a plasma. Only
part of the electric field $E_{t}(x)$ is shown. Solid lines show the exact
solution given by Eq.~(\protect\ref{Et profile text}); dashed lines and
subscript $appr$ correspond to the approximate calculation of Eq.~(  \protect
\ref{Et profile text appr}); dotted lines and the subscript $std$ correspond
to the approximate calculation in the limit $x\gg \protect\lambda _{\protect%
\omega }$ obtained making use the method of the steepest descend given by
Eq.~(\protect\ref{Et landau text}). The rf electric field frequency is 13.56
MHz and the plasma density is 10$^{8}$cm$^{-3}$ (lines) and 10$^{9}$cm$^{-3}$
(symbols).}
\label{Figure Et}
\end{figure}

To summarize, the solution can be separated into three parts,
\begin{equation}
E_{x}(x,t)=\left[ E_{0}\exp \left( -\frac{x}{a}\right) +E_{b}+E_{t}(x)\right]
e^{-i\omega t}.  \label{Efield profile text}
\end{equation}%
Here, $E_{0}$ is the amplitude of the electric field at the plasma boundary,
$E_{b}=E_{0}/\varepsilon $ is the electric field in the plasma bulk far away
from the sheath region, $\varepsilon =1-\omega _{p}^{2}/\omega ^{2}$ is the
dielectric constant of the cold plasma, and $E_{t}(x)$ is the electric field
in a transient region with a spatial length of order $v_{T}/\omega $. The
first term is the Debye screening of the external electric field. The second
part describes a small, uniform electric field penetrating into the plasma
far away from the boundary. The second and third terms are absent for a
stationary applied electric field and appear only in the case of the rf
electric field. The solution for the transient electric field $E_{t}(x)$
profile is derived in Appendix B and is given by
\begin{equation}
E_{t}(x)=\frac{2E_{0}}{\pi }\int_{0}^{\infty }\frac{1}{k}\frac{%
Im[\varepsilon _{\Vert }(\omega ,k)]}{\varepsilon _{\Vert }^{\ast }(\omega
,k)\;\varepsilon _{\Vert }(\omega ,k)}e^{ikx}dk,  \label{Et profile text}
\end{equation}%
where $\varepsilon _{\Vert }(\omega ,k)$ is the longitudinal plasma
permittivity ($\mathbf{E}\Vert \mathbf{k}$),
\begin{equation}
\varepsilon _{\Vert }(\omega ,k)\simeq 1+\frac{2\omega _{p}^{2}}{%
k^{2}v_{T}^{2}}\left[ 1+\frac{\omega }{kV_{T}}Z\left( \frac{\omega }{kV_{T}}%
\right) \right] ,  \label{dielectric function 4}
\end{equation}%
and $Z(\zeta )$ is the plasma dispersion function \cite{Plasma Formulary}
\begin{equation}
Z(\xi )=\pi ^{-1/2}\int_{-\infty }^{\infty }dt\frac{\exp \left(
-t^{2}\right) }{t-\xi },\;\;\;Im\xi >0.  \label{PDF}
\end{equation}%
In the limit $x\gg \lambda _{\omega }$ only small $k$ contribute to the
integral and $\varepsilon _{\Vert }(\omega ,k)$ can be substituted by $%
\varepsilon _{\Vert }^{\ast }(\omega ,0)\equiv \varepsilon $ in the
denominator of Eq.~(\ref{Et profile text appr}), which gives
\begin{equation}
E_{t}(x)\approx E_{appr}(x)=\frac{2E_{0}}{\pi \varepsilon ^{2}}%
\int_{0}^{\infty }\frac{1}{k}Im[\varepsilon _{\Vert }(\omega ,k)]e^{ikx}dk,
\label{Et profile text appr}
\end{equation}%
Application of the method of steepest descend to Eq.~(\ref{Et profile text
appr}) yields \cite{Landau}
\begin{equation}
E_{t}(x)\approx E_{std}(x)=\frac{2E_{0}}{\sqrt{3}\varepsilon ^{2}}\frac{%
\omega _{p}^{2}}{\omega ^{2}}\left( \frac{x}{\lambda _{\omega }}\right)
^{2/3}\exp \left[ \left( -\frac{3}{4}+i\frac{3\sqrt{3}}{4}\right) \left(
\frac{x}{\lambda _{\omega }}\right) ^{2/3}-i\pi /3\right] ,
\label{Et landau text}
\end{equation}%
where $\lambda _{\omega }=v_{T}/\sqrt{2}\omega $ is the phase-mixing scale.
The plots of the amplitude and phase of the electric field profile $E_{t}(x)$
given by Eq.(\ref{Et profile text}) and the approximate $E_{appr}(x)$ given
by Eq.~(\ref{Et profile text appr}), and asymptotic analytical result $%
E_{std}(x)$ given by Eq.~(\ref{Et landau text}) are shown in Fig.~\ref%
{Figure Et}. Figure \ref{Figure Et} shows that the steepest descend method
given by Eq.~(\ref{Et landau text}) closely approximates Eq.~(\ref{Et
profile text appr}) already for $x>v_{T}/\omega $. However, the both
asymptotic solutions in Eq.~(\ref{Et profile text appr}) and Eq.(\ref{Et
landau text}) approximate the full solution in Eq. (\ref{Et profile text})
only for very large $x>40V_{T}/\omega $. This is due to the made
substitution $\varepsilon (\omega ,k)$ by $\varepsilon (\omega ,0)$, which
results in a considerable error for $k\sim \omega /v_{T}$ or $x\sim
v_{T}/\omega $.

\emph{It follows from Eq.~(\ref{Et landau text}) that the electric field
amplitude at $x>v_{T}/\omega $\ is of order $E_{0}/\varepsilon $, i.e., it
is comparable with the electric field far away from the boundary (}$%
E_{t}\sim E_{b}$\emph{)}.\emph{\ }

The origin of the electric field $E_{t}(x)$ can be explained by analyzing
the individual electron dynamics. After passing through the region of the rf
field, an electron acquires changes $\Delta \varepsilon (v_{x})$ in energy
and $\Delta u(v_{x})$ in velocity
\begin{eqnarray}
\Delta \varepsilon (v_{x}) &=&\int_{-\infty }^{\infty }v_{x}eE[x(t),t]dt,
\label{u kick} \\
\Delta u(v_{x}) &=&\frac{\Delta \varepsilon }{mv_{x}}.  \nonumber
\end{eqnarray}%
Here, the electron trajectory is $x(t)=v_{x}t$, $v_{x}=|v_{x}|sgn(t)$, and
the electric field profile is given by Eq.~(\ref{Efield profile text}). The
total velocity kick is the summation over velocity kicks due to exponential,
bulk and transitional electric fields%
\begin{equation}
\Delta u(v_{x})=\Delta u_{0}+\Delta u_{b}+\Delta u_{t}.
\end{equation}%
Substituting an exponential electric field into Eq.(\ref{u kick}) gives the
corresponding electron velocity kick
\begin{equation}
\Delta u_{0}(v_{x})\simeq \frac{2eE_{0}}{m\omega }\frac{\omega ^{2}a^{2}}{%
v_{x}^{2}+(\omega a)^{2}}.  \label{d v a}
\end{equation}%
Substituting the uniform electric field $E_{b}$ into Eq.~(\ref{u kick}),
gives the electron velocity
\begin{equation}
\Delta u_{b}(v_{x})\simeq \frac{2eE_{0}}{im\omega \varepsilon }.
\label{dv b}
\end{equation}%
This calculation can also be explained as follows: An electron has the
oscillating velocity $\Delta u_{s}=eE_{b}i/m\omega $ in a uniform rf
electric field and a thermal velocity $v_{x}$. After a collision with the
wall, an electron changes its velocity direction. If the initial average
velocity was $v_{x}<0$, after the collision with the wall with specular
reflection, the new average velocity $v_{x}^{\prime }>0$ will change
according to
\begin{equation}
v_{x}^{\prime }+\Delta u_{s}(t)=-[v_{x}+\Delta u_{s}(t)]
\end{equation}%
or the average velocity changes to%
\begin{equation}
v_{x}^{\prime }=-v_{x}-2\Delta u_{s}(t),
\end{equation}%
which results in the effective velocity kick of Eq.~(\ref{dv b}).

The origin of the electric field in the transition region $E_{t}(x)$ is due
to the plasma polarization. The velocity perturbations $\Delta
u_{s}(v_{x},t) $ produce bunches in the electron density, which, in turn,
generate the electric field $E_{t}(x)$. The decay of the electric field $%
E_{t}(x)$ is due to phase mixing similarly to the test-particle case in Eq.~(%
\ref{current approx}). Thus, generation of the transitional electric field $%
E_{t}(x)$ can be considered as a plasma self-consistency effect.

\begin{figure}[th]
\begin{center}
\includegraphics[width=3in]{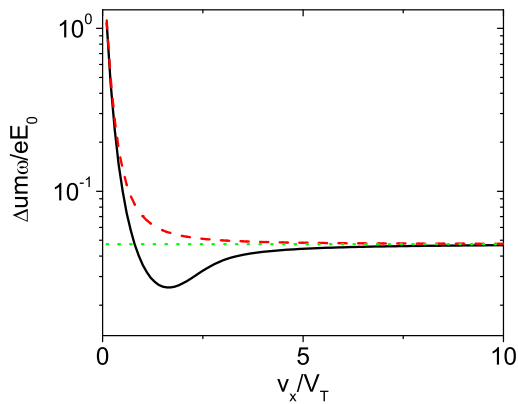}
\end{center}
\caption{{}Electron velocity kick after interaction with the rf electric
field. Solid line shows a velocity kick $\Delta u$ calculated according to
the full electric field in Eq.~(\protect\ref{Efield profile text}). The
dashed line shows a $\Delta u$ due to the electric field $E_{0}\exp (-x/a)$
and the uniform electric field $E_{b}$ only; the dotted line is due to $%
E_{b} $. The rf electric field frequency is 13.56 MHz and the plasma density
is 10$^{8}$cm$^{-3}$. }
\label{Figure Vkick}
\end{figure}

The electric field $E_{t}(x)$ generates a significant portion of the total
velocity kick and thus noticeably influences the electron heating in the rf
electric field. Figure \ref{Figure Vkick} shows the amplitude of the
electron velocity kick $\Delta u(v_{x})$ due to the interaction with the
electric field given by Eq.~(\ref{Efield profile text}). Electrons with
small velocities $v_{x}\sim \omega a=v_{T}\omega /\omega _{p}$ pick up a
large velocity kick due to the exponential electric field $E_{0}\exp
(-x/a-i\omega t)$, $\Delta u\simeq \Delta u_{0}\sim 2eE_{0}/m\omega $. For
very large electron velocities $v_{x}\gg v_{T}$, the velocity kick given by
Eq.~(\ref{d v a}) becomes small and the main contribution to the velocity
kick comes from the uniform electric field $E_{b}=E_{0}e^{-i\omega
t}/\varepsilon $ and the collision with the wall, $\Delta u\simeq \Delta
u_{b}\sim 2eE_{0}/m\omega \varepsilon $. In the intermediate range of
velocities $v_{x}\sim v_{T}$, the account of the electric field $E_{t}(x)$
is important, as in this case $\Delta u_{t}\sim \Delta u_{b}$. As is evident
from Fig.~\ref{Figure Vkick}, taking this electric field $E_{t}(x)$ into
account results in a considerable reduction of the electron velocity kick $%
v_{x}\sim v_{T}$ for the bulk of the electron population compared with the
case when this electric field is not taken into account. Note that most
models neglect the electric field $E_{t}(x)$, see for example \cite%
{Libermann89},\cite{Lieberman& Godyak review}.

\subsection{Large amplitude electric field}

In many practical applications, the value of the external electric field is
large: the potential drop in the sheath region $V_{sh}$ is typically of the
order of hundreds of Volts and is much larger than the electron temperature $%
T_{e}$, which is of the order of a few Volts; consequently the electric
field penetration has to be treated nonlinearly.

In the limit $V_{sh}\gg T_{e}$, a wall is charged negatively all time with
an alternating charge in a manner to conduct an ac current, driven by an
external electric circuit. A negative charge pushes electrons away from the
electrode up to a distance where the negative electric field is screened by
a positive ion density. As $V_{sh}\gg T_{e}$, the sheath width is much
larger than the Debye length and the plasma sheath boundary can be
considered as infinitely thin. The position of the boundary is determined by
the condition that the external electric field is screened in the sheath
regions when and where electrons are absent \cite{Libermann89, Me and
Tsendin 1992 1}.

\begin{figure}[tbh]
\centering\includegraphics[width=3in]{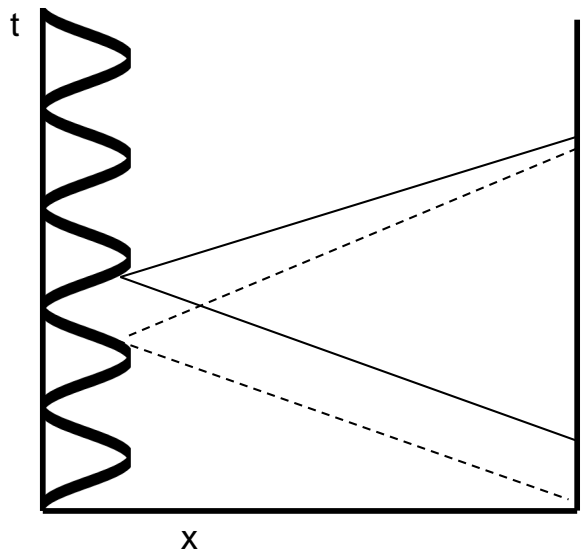}
\caption{Schematic of density bunches formation in the region adjacent to
the sheath. The plasma-sheath boundary is shown by bold solid line.
Electrons with the same velocity $v_{x}$ and distance apart $\sim v_{x}/%
\protect\omega $ collide with the sheath. The first electron looses its
energy and decelerates, whereas the second acquire energy and accelerates.
As a result, the distance between two electrons decreases, which produces
electron density perturbations. }
\label{CCPbunches}
\end{figure}

Electron interactions with the sheath electric field are traditionally
treated as collisions with a moving potential barrier (wall). It is well
known that multiple electron collisions with an oscillating wall result in
electron heating, provided there is sufficient phase-space randomization in
the plasma bulk. It is common to describe the sheath heating by considering
electrons as test particles, and neglecting the plasma electric field \cite%
{Lieberman& Godyak review}. As was pointed out in Refs.~\cite{Libermann89,
Me and Tsendin 1992 2, Aliev and me} accounting for the electric field in
the plasma reduces the electron sheath heating, and the electron sheath
heating vanishes completely in the limit of uniform plasma density.
Therefore, an accurate description of the rf fields in the bulk of the
plasma is necessary for calculating the sheath heating. The electron
velocity is oscillatory in the sheath, and as a result of these velocity
modulations, the electron density bunches appear in the region adjacent to
the sheath, similar to the previously described case of small-amplitude
wave, see Fig.~\ref{CCPbunches}. These electron density perturbations decay
due to phase mixing over a length of order $v_{T}/\omega ,$ where $v_{T}$ is
the electron thermal velocity, and $\omega $\ is the frequency of the
electric field. The electron density perturbations polarize the plasma and
produce an electric field in the plasma bulk. This electric field, in turn,
changes the velocity modulations and correspondingly influences the electron
density perturbations. Therefore, electron sheath heating has to be studied
in a self-consistent nonlocal manner assuming a finite-temperature plasma.

Notwithstanding the fact that particle-in-cell simulations results have been
widely available for the past decade \cite{Sommerer, Surendra PRL}, a basic
understanding of the electron heating by the sheath electric field is being
incomplete, because no one has studied the electric field in the plasma bulk
using a kinetic approach, similar to the anomalous skin effect for the
inductive electric field \cite{Lifshitz and Pitaevskii}. In this regard,
analytical models are of great importance because they shed light on the
most complicated features of collisionless electron interactions with the
sheath. In Ref.\cite{My PRL 2002}, an analytical model was developed to
explore the effects associated with the self-consistent non-local nature of
this phenomenon.

One of the approaches to study electron sheath heating is based on a fluid
description of the electron dynamics. For the collisionless case, closure
assumptions for the viscosity and heat fluxes are necessary. In most cases,
the closure assumptions are made empirically or phenomenologically \cite%
{Surendra PRL}, \cite{Gozadinos}. The closure assumptions have to be
justified by direct comparison with the results of kinetic calculations as
is done, for example, in Refs. \cite{Hammett,Furkal}. Otherwise, inaccurate
closure assumptions may lead to misleading results as discussed below.
\begin{figure}[tbh]
\centering\includegraphics[width=65mm]{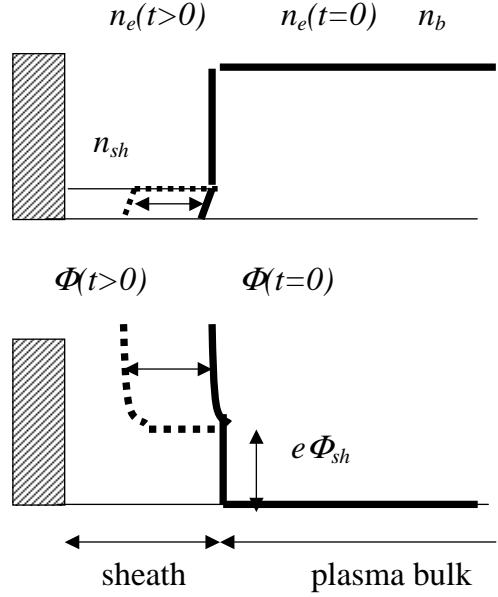}
\caption{Schematic of a sheath. The negatively charged electrode pushes
electrons away at different distances depending on the strength of the
electric field at the electrode. Shown are the density and potential
profiles at two different times. The solid line is at the time of maximum
sheath expansion.}
\label{CCPscheme}
\end{figure}
To model the sheath-plasma interaction analytically, the following
simplifying assumptions have been adopted in Ref.~\cite{My PRL 2002}. The
discharge frequency is assumed to be small compared with the electron plasma
frequency. Therefore, most of the external electric field is screened in the
sheath region by an ion space charge. The ion response time is typically
larger than the inverse discharge frequency, and the ion density profile is
quasi-stationary. There is an ion flow from the plasma bulk towards the
electrodes. In the sheath region, ions are being accelerated towards the
electrode by the large sheath electric field, and the ion density in the
sheath region is small compared with the bulk ion density. In the present
analytical treatment, the ion density profile is assumed fixed and is
modelled in a two-step approximation: the ion density $n_{b}$ is uniform in
the plasma bulk, and the ion density in the sheath $n_{sh}<n_{b}$ is also
uniform (see Fig.~\ref{CCPscheme}). At the sheath-plasma boundary, there is
a stationary potential barrier for the electrons ($e\Phi _{sh}$), so that
only the energetic electrons reach the sheath region. The potential barrier
is determined by the quasineutrality condition, i.e., when the energetic
electrons enter the sheath region, their instantaneous density is equal to
the ion density [$n_{e}(\Phi _{sh})=n_{sh}$].

The electron density profile is time-dependent in response to the
time-varying sheath electric field. The large sheath electric field does not
penetrate into the plasma bulk. Therefore, the quasineutrality condition
holds in the plasma bulk, i.e., the electron density is equal to ion
density, $n_{e}=n_{b}.$ In the sheath region, the electrons are reflected by
the large sheath electric field. Therefore, $n_{e}=n_{sh}$ for $x>x_{sh}(t)$%
, and $n_{e}=0$ for $x<x_{sh}(t)$,\ where $x_{sh}(t)$ is the position of the
plasma-sheath boundary \cite{Libermann89}. From Maxwell's equations it
follows that $\mathbf{\nabla \cdot J}=0$, where the total current $\mathbf{J}
$ is the sum of the displacement current and the electron current. In the
one-dimensional case, the condition $\mathbf{\nabla \cdot J}=0$ yields the
conservation of the total current \cite{Landau, Me and Tsendin 1992 1}:
\begin{equation}
en_{e}V_{e}+\frac{1}{4\pi }\frac{\partial E_{sh}}{\partial t}=j_{0}\sin
(\omega t+\phi ),  \label{Esh}
\end{equation}%
where $j_{0}$ is the amplitude of the rf current controlled by an external
circuit and $\phi $ is the initial phase. In the sheath, electrons are
absent in the region of large electric field, and Eq.(\ref{Esh}) can be
integrated to give \cite{Me and Tsendin 1992 1}
\begin{equation}
E_{sh}(x,t)=\frac{4\pi j_{0}}{\omega }[-1-\cos (\omega t+\phi )]+4\pi
|e|n_{sh}x,\quad x<x_{sh}(t)  \label{Esh(x,t)}
\end{equation}%
where Poisson's equation has been used to determined the spatial dependence
of the sheath electric field. The first term on the right-hand side of Eq.~(%
\ref{Esh(x,t)}) describes the electric field at the electrode and the second
term relates to the ion space charge screening of the sheath electric field.
The position of the plasma-sheath boundary $x_{sh}(t)$ is determined by the
zero of the sheath electric field, $E_{sh}[x_{sh}(t),t]=0$. From Eq.~(\ref%
{Esh(x,t)}) it follows that
\begin{equation}
x_{sh}(t)=\frac{V_{sh0}}{\omega }[1+\cos (\omega t+\phi )],  \label{xsh(t)}
\end{equation}%
where $V_{sh0}=j_{0}/(en_{sh})$ is the amplitude of the plasma-sheath
boundary velocity. The ion flux on the electrode is small compared with the
electron thermal flux. Because electrons attach to the electrode, the
electrode surface charges negatively, so that in a steady-state discharge,
the electric field at the electrode is always negative, preventing an
electron flux on the electrode. However, for a very short time ($\omega
t_{n}+\phi \approx \pi (1+2n)$) the sheath electric field vanishes, allowing
electrons to flow to the electrode for compensation of the ion flux. Note
that there is a large difference between the sheath structure in the
discharge and the sheath for obliquely incident waves interacting with a
plasma slab without any bounding walls. Because electrodes are absent,
electrons can move outside the plasma, and the electric field in the vacuum
region, $E_{sh}(x,t)=(4\pi j_{0}/\omega )\cos (\omega t+\phi )$, may have an
alternating sign. Therefore, electrons may penetrate into the region of
large electric field during the time when $E_{sh}(x,t)>0$ \cite{Brunel, Yang}%
. In the discharge, however, because the sheath electric field given by Eq.~(%
\ref{Esh(x,t)}) always reflects electrons, the electrons \emph{never} enter
the region of the large sheath electric field, which is opposite to the case
of obliquely incident waves.

The calculations based on the two-step ion density profile model are known
to yield discharge characteristics in good agreement with experimental data
and full-scale simulations \cite{Orlov}.

For analytical calculation of the rf electric field inside the plasma, a
linear approximation is used for the plasma conductivity. The validity of
the linear approximation is based on the fact that the plasma-sheath
boundary velocity and the mean electron flow velocity are small compared
with the electron thermal velocity, $V_{sh}\ll v_{T}$, \cite{Me and Tsendin
1992 1, Sommerer}. The important spatial scale is the length scale for phase
mixing, $\lambda _{\omega }$. The sheath width satisfies $2V_{sh0}/\omega
\ll \lambda _{\omega }$ because $V_{sh}\ll v_{T}$. Therefore, the sheath
width is neglected, and electron interactions with the sheath electric field
are treated as a boundary condition. The collision frequency ($\nu $) is
assumed to be small compared with the discharge frequency ($\nu \ll \omega $%
), and correspondingly the mean free path is much larger than the length
scale for phase mixing. Therefore, the electron dynamics is assumed to be
collisionless. The discharge gap is considered to be sufficiently large
compared with the electron mean free path, so that the influence of the
opposite sheath is neglected. The effects of a finite gap width have been
discussed in Refs. \cite{Me PRL 1999,Ulrich and me}.

The electron interaction with the large electric field in the sheath is
modelled as a collision with a moving oscillating rigid barrier with
velocity $V_{sh}(t)=dx_{sh}(t)/dt$. After a collision with the plasma-sheath
boundary - modelled as a rigid barrier moving with velocity $V_{sh}(t)$ - an
electron with initial velocity $-u$ acquires a velocity $u+2V_{sh}$.
Therefore, the power deposition density transfer from the oscillating
plasma-sheath boundary is given by \cite{Libermann89}

\begin{equation}
P_{sh}=\frac{m}{2}\left\langle \int_{-V_{sh}}^{\infty }du\left[ u+V_{sh}(t) %
\right] \left[ (2V_{sh}(t)+u)^{2}-u^{2}\right] \,f_{sh}(-u,t)\right\rangle ,
\label{Psh}
\end{equation}%
where $m$ is the electron mass, $f_{sh}(-u,t)$ is the electron velocity
distribution function in the sheath, and $\left\langle \cdot \cdot \cdot
\right\rangle $ denotes a time average over the discharge period.
Introducing a new velocity distribution function $g(-u^{\prime
},t)=f_{sh}[-u-V_{sh}(t),t]$, Eq.~(\ref{Psh}) yields

\begin{equation}
P_{sh}=-2m\left\langle V_{sh}(t)\int_{0}^{\infty }u^{\prime 2}g(-u^{\prime
},t)du^{\prime }\right\rangle ,  \label{Psh
modified}
\end{equation}%
where $-u^{\prime }=-u-V_{sh}$ is the electron velocity relative to the
oscillating rigid barrier. From Eq.(\ref{Psh modified}) it follows that, if
the function $g(u^{\prime })$ is stationary, then ($P_{sh}=0$) and there is
no collisionless power deposition due to electron interaction with the
sheath \cite{Libermann89, Gozadinos, Raizer book}. For example, in the limit
of a uniform ion density profile $n_{sh}=n_{b}$, $g(u^{\prime })$ is
stationary (\emph{in an oscillating reference frame of the plasma-sheath
boundary}), and the electron heating vanishes \cite{Libermann89}, \cite{Me
and Tsendin 1992 1}. Indeed, in the plasma bulk, the displacement current is
small compared with the electron current, and from Eq.~(\ref{Esh}) it
follows that the electron mean flow velocity in the plasma bulk, $%
V_{b}(t)=-j_{0}\sin (\omega t+\phi )/en_{b}$, is equal to the plasma-sheath
velocity $V_{sh}(t)$, from Eq.~(\ref{xsh(t)}). \emph{Therefore, the electron
motion in the plasma is strongly correlated with the plasma-sheath boundary
motion.} From the electron momentum equation it follows that there is an
electric field, $E_{b}=m/e\,dV_{b}(t)/dt$, in the plasma bulk.\ In a frame
of reference moving with the electron mean flow velocity, the sheath barrier
is stationary, and there is no force acting on the electrons, because the
electric field is compensated by the inertial force ($%
eE_{b}-mdV_{b}(t)/dt=0) $. Therefore, electron interaction with the sheath
electric field is totally compensated by the influence of the bulk electric
field, and the collisionless heating vanishes \cite{Me and Tsendin 1992 2}.
The example of a uniform density profile shows the importance of a
self-consistent treatment of the collisionless heating in the plasma. If the
function $g(u^{\prime },t)$ is nonstationary, there is net power deposition.
In Ref. \cite{My PRL 2002}, a kinetic calculation is performed to yield the
correct electron velocity distribution function $g(u^{\prime },t)$ and,
correspondingly, the net power deposition.

The electron motion is different for low-energy electrons with an initial
velocity in the plasma bulk $|u|<u_{sh}$, where $u_{sh}^{2}=2e\Phi _{sh}/m$
and for energetic electrons with velocity $|u|>u_{sh}$. The low energy
electrons with initial velocity $-u$ in the plasma bulk are reflected from
the stationary potential barrier $e\Phi _{sh}$, and then return to the
plasma bulk with velocity $u$. High energy electrons enter the sheath region
with velocity $u_{1}=-(u^{2}-u_{sh}^{2})^{1/2}$. They acquire a velocity $%
u_{2}=2V_{sh}-u_{1}$ after collision with the moving rigid barrier, and then
return to the plasma bulk with a velocity $(u_{2}^{2}+u_{sh}^{2})^{1/2}$
\cite{multiple collisions}.

As the electron velocity is modulated in time during reflections from the
plasma-sheath boundary, so is the energetic electron density (by continuity
of the electron flux). This phenomenon is identical to the mechanism of
klystron operation \cite{klystron}. The perturbations in the energetic
electron density yield an electric field in the transition region adjusted
to the sheath, see Fig.\ref{CCPbunches}.

The solution for the electric field $E_{t}(x)$ was obtained analytically in
Ref.\cite{My PRL 2002}. Similar to the previous section, the solution is an
expression for the inverse Fourier transform. It cannot be represented in an
analytical form and has to be simulated numerically. This simulation has
been performed for $n_{sh}/n_{b}=1/3$, $\omega /\omega _{p}=1/100$, and a
Maxwellian electron distribution function. The electric field profile is
close to $E_{t}(x)\approx E_{t0}\exp (-x/\lambda _{c})$, where $%
E_{t0}=-0.72T_{e}/\lambda _{\omega }$, and $\lambda _{c}=(0.19+0.77i)\lambda
_{\omega }$ for $x<6V_{T}/\omega $. For $x>6V_{T}/\omega $, the electric
field profile is no longer a simple exponential function, which is similar
to the case considered in the previous section. The difference in phase of
the currents of the energetic and low-energy electrons was observed in Ref.%
\cite{Surendra PRL}, but it was misinterpreted as the generation of the
electron acoustic waves. Electron acoustic waves can be excited if there is
a complex value of $k$, with small damping $\mathit{{Im}(k)\ll {Re}(k),}$
which is the root of the plasma dielectric function $\varepsilon (\omega
,k)=0$ for a given $\omega $. For a Maxwellian electron distribution
function, such root does not exist when $\omega \ll \omega _{p}$. However,
the electron acoustic waves can exist if the plasma contains two groups of
electrons which have very different temperatures \cite{Mace}. The wave phase
velocity is $\omega /k=\sqrt{n_{c}/n_{h}}\sqrt{T_{h}/m}$ , where $n_{c}$ and
$n_{h}$ are the electron densities of cold and hot electrons, respectively,
and $T_{h}$ is the temperature of the hot electrons. The electron acoustic
waves are strongly damped by the hot electrons, unless $n_{c}\ll n_{h}$ and $%
T_{c}\ll T_{h}$ , where $T_{c}$ is the electron temperature of the cold
electrons \cite{Mace}. In the opposite limit, $n_{c}>4n_{h}$, the electron
acoustic waves do not exist \cite{Mace}. In capacitively-coupled discharges,
the electron population does stratify into two populations of cold and hot
electrons, as has been observed in experiments \cite{Godyak 2EDF} and
simulation studies \cite{cold electron formation, My CCP}. Cold electrons
trapped by the plasma potential in the discharge center do not interact with
the large electric fields in the sheath region and have low temperature.
Moreover, because of the nonlinear evolution of plasma profiles, the cold
electron density is much larger than the hot electron density \cite{cold
electron formation}. Therefore, weakly-damped electron acoustic waves do not
exist in the plasma of capacitively-coupled discharges. Reference \cite%
{Surendra PRL}\ used the fluid equation and neglected the effect of
collisionless dissipation, thus arriving at the incorrect conclusion about
the existence of weakly-damped electron acoustic waves.
\begin{figure}[tbh]
\includegraphics[width=3in]{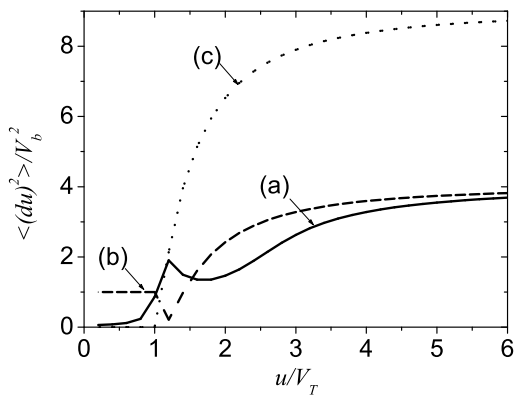}
\caption{Plot of the averaged square of the dimensionless velocity kick as a
function of the dimensionless velocity for the conditions in Fig.1, taking
into account (a) both $E_{1}(x)$ and $E_{b}$ - solid line; (b) only $E_{b}$
- dashed line; and (c) no electric field - dotted line.}
\label{average2velocitykick}
\end{figure}

The power deposition is given by the sum of the power transferred to the
electrons by the oscillating rigid barrier in the sheath region and by the
electric field in the transition region,
\begin{equation}
P_{tot}=P_{sh}+P_{tr}.  \label{Power tot}
\end{equation}%
Note that $P_{tr}$ can be negative. Calculations making use of the Vlasov
equation yield \cite{My PRL 2002}
\begin{equation}
P_{tot}=-\int_{0}^{\infty }muD_{u}(u)\frac{df_{0}}{du}du,  \label{Ptot}
\end{equation}%
where
\begin{equation}
D_{u}(u)=\frac{u|du|^{2}}{4}  \label{D(u)}
\end{equation}%
is the diffusion coefficient in velocity space, and $du$ is the change in
the electron velocity after passing through the transition and sheath
regions,
\begin{equation}
du=2iV_{b}\left[ \frac{u^{\prime }}{u}\frac{n_{b}}{n_{sh}}\Theta
(|u|-u_{sh})-1\right] +\frac{eE_{t}(k=\omega /u)}{u},  \label{du}
\end{equation}%
where $E_{t}(k)$ is the Fourier transform of the electric field $E_{t}(x)$.
First term describes the velocity acquired by fast electrons ($|u|>u_{sh}$)
in collisions with the sheath; the second is due to the bulk electric field $%
E_{b}$ and collisions with either the potential barrier $\Phi _{sh}$ or
sheath; and the third is due the electric field in the transitional region $%
E_{t}(x)$. A plot of $|du|^{2}/4$ is shown in Fig.~\ref{average2velocitykick}%
. Taking into account the electric field in the plasma (both $E_{b}$ and $%
E_{t}$) reduces $|du|$ for energetic electrons ($u>u_{sh}$) and increases $%
|du|$ for slow electrons ($u<u_{sh}$). \emph{Therefore, the electric field
in the plasma cools the energetic electrons and heats the low-energy
electrons, respectively.} Similar observations were made in numerical
simulations \cite{Surendra PRL}.
\begin{figure}[tbh]
\includegraphics[width=3in]{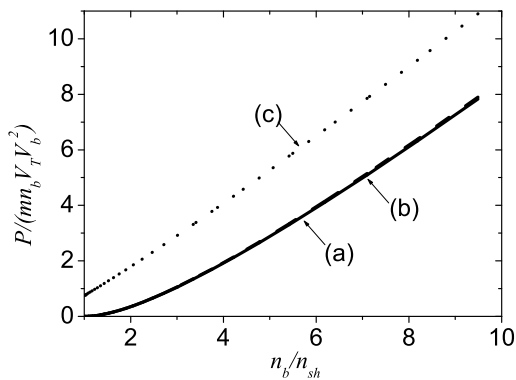}
\caption{Plot of the dimensionless power density as a function of the ratio
of the bulk plasma density to the sheath density, taking into account (a)
both $E_{1}(x)$ and $E_{b}$ - solid line; (b) only $E_{b}$ - dashed line;
and (c) no electric field inside the plasma - dotted line. }
\label{PowerCCP}
\end{figure}
Figure \ref{PowerCCP} shows the dimensionless power density as a function of
$n_{b}/n_{sh}$. Taking into account the electric field in the plasma (both $%
E_{b}$ and $E_{1}$) reduces the total power deposited in the sheath region.
Interestingly, taking into account only the uniform electric field $E_{b}$
gives a result close to the case when both $E_{b}$ and $E_{1}$ are accounted
for. The electric field $E_{1}$ redistributes the power deposition from the
energetic electrons to the low energy electrons, but does not change the
total power deposition (compare lines (a) and (b) in Fig.\ref%
{average2velocitykick} and Fig.~\ref{PowerCCP}). Therefore, the total power
deposition due to sheath heating can be calculated approximately from Eq.~(%
\ref{Ptot}), taking into account only the electric field $E_{b}$. This gives
\begin{equation}
P_{tot}\approx -mV_{b}^{2}\int_{0}^{\infty }u^{2}\left[ \frac{u^{\prime }}{u}%
\frac{n_{b}}{n_{sh}}\Theta (u-u_{sh})-1\right] ^{2}\frac{df_{0}}{du}du.\,
\label{Ptotfinal}
\end{equation}%
The result of the self-consistent calculation of the power dissipation in
Eq.~(\ref{Ptotfinal}) differs from the non-self-consistent estimate by the
last term in Eq.~(\ref{Ptotfinal}), which contributes corrections of order $%
n_{sh}/n_{b}$ to the main term.

A future development should provide a self-consistent analysis of a more
realistic, nonuniform, and self-consistent ion density profile $n_{i}(x)$.
Such study has been currently performed for inductively coupled discharges
only.

\section{Penetration of the rf electric field into an inductively-coupled
plasma}

Low pressure inductively-coupled rf discharges are often operated in the
non-propagating regime, when the driving rf field penetrates into plasma
only within a skin layer of width $\delta $ near the antenna, i.e., exhibits
a skin effect. Not only the rf field, but, in this case, also the resulting
induced electric current is concentrated near the surface of the plasma.
Depending on the local, or non-local nature of the relation between the
electric current $j$ induced in plasma and the rf electric field $E$, the
skin effect is called \textit{normal}, if the dependence of the current on
the electric field is local, or \textit{anomalous}, if the dependence of the
current on the electric field is nonlocal \cite{Kolobov review}.

To differentiate between the two regimes of the skin effect, it is
convenient to introduce the nonlocality parameter \cite{Weibel} $\Lambda
=(\lambda /\delta _{0})^{2}$, where $\lambda \equiv v_{T}/(\omega ^{2}+\nu
^{2})^{1/2}$ is the effective electron mean free path and
\begin{equation}
\delta _{0}=\frac{c}{\omega _{p}(1+\nu ^{2}/\omega ^{2})^{1/4}}
\label{normal skin effect}
\end{equation}%
is the depth of the normal skin effect. The parameter $\Lambda $%
\begin{equation}
\Lambda =\frac{v_{T}^{2}\omega _{p}^{2}}{\omega ^{2}c^{2}(1+\nu ^{2}/\omega
^{2})^{1/2}}
\end{equation}%
is a fundamental measure of plasma current non-locality. In the local limit $%
\Lambda \ll 1$, the effective mean free path is small compared with the skin
depth $\lambda \ll \delta $, and the current density at a particular point
in space can be considered as a function of the electric field at the same
point $\mathbf{j}(\mathbf{x})=\sigma (\mathbf{x})\mathbf{E}(\mathbf{x})$
(Ohm's law). In the opposite limit $\lambda \gg \delta $, the mean free path
exceeds the skin depth $\lambda \gg \delta $, the relation between the
current and the field $\mathbf{j}(x)=\int \underline{\sigma }(\mathbf{x},%
\mathbf{x}^{\prime })\mathbf{E}(\mathbf{x}^{\prime })d\mathbf{x}^{\prime }$
is no longer local, because the conductivity $\sigma (\mathbf{x},\mathbf{x}%
^{\prime })$ has a spatial dispersion.

The penetration of the rf electric field into the plasma is described
according to Faraday's and Ampere's laws
\begin{equation}
\mathbf{\nabla }\times \mathbf{E}=-\frac{1}{c}\frac{\partial \mathbf{B}}{%
\partial t},  \label{Faraday's}
\end{equation}%
\begin{equation}
\mathbf{\nabla }\times \mathbf{B}=\frac{1}{c}\frac{\partial \mathbf{D}}{%
\partial t}+\frac{4\pi }{c}\mathbf{j}.  \label{Ampere's}
\end{equation}

For a transverse harmonic wave in one-dimensional geometry $%
E_{y}(x)e^{-i\omega t}$, the Faraday's and Ampere's laws give
\begin{equation}
\left( \frac{\partial ^{2}}{\partial x^{2}}+\frac{\omega ^{2}}{c^{2}}\right)
E_{y}\mathbf{=-}\frac{4\pi i\omega }{c^{2}}j_{y},  \label{Maxwell 1D}
\end{equation}%
where the current $j$ is the plasma electron current $j_{y}=j_{ey}$ (the
ions are considered stationary), which has to be calculated making use of
the electron kinetic equation, similar to the case of the penetration of the
longitudinal wave into the plasma described in the previous section.

\subsection{Normal skin effect}

In the limit of the normal skin effect ($\Lambda \ll 1$), the electron
thermal motion can be neglected. The electron flow velocity $V_{ey}$ may be
obtained from Newton's law taking into account the drag force due to the
electron neutral collisions,
\begin{equation}
m\frac{\partial }{\partial t}V_{ey}=-eE_{y}-\nu V_{ey}.
\end{equation}%
This gives for the electron current ($j_{ey}=-en_{e}V_{ey})$ the Ohm's law
relationship
\begin{equation}
\mathbf{j}_{e}(x)=\sigma _{e}\mathbf{E}(x),  \label{Ohm's law}
\end{equation}%
where%
\begin{equation}
\sigma _{e}=\frac{e^{2}n_{e}}{m(\nu -i\omega )}.  \label{sigma1}
\end{equation}%
The plasma current density is proportional to the electric field at the same
point of space with a proportionality coefficient that is the complex
conductivity of the cold plasma. Substituting Ohm's law Eq.~(\ref{Ohm's law}%
) with plasma conductivity from Eq.~(\ref{sigma1}) into Eq.~(\ref{Maxwell 1D}%
) gives the solution of the wave equation
\begin{equation}
E_{y}=E_{y0}e^{-\alpha x},
\end{equation}%
where $\alpha =\sqrt{-4\pi i\omega \sigma _{e}/c^{2}}$. Here, we neglected
small terms associated with the displacement current in the limit $\omega
\ll \omega _{p}$, which is valid for the most plasma parameters in ICP
discharges. The electric field can be equivalently expressed as
\begin{equation}
E_{y}(x,t)=E_{y0}e^{-cos(\epsilon /2)x/\delta _{0}}\cos [\omega t-\sin
(\epsilon /2)x/\delta _{0}],  \label{normal skin law}
\end{equation}%
where $\delta _{0}$ is the normal skin depth in Eq.(\ref{normal skin effect}
), and $\epsilon =\arctan (\nu /\omega )$.

\subsection{Anomalous skin effect}

The case of anomalous skin effect ($\Lambda \geq 1$) for low-pressure
inductively-coupled plasmas is more complicated comparing to the case of
normal skin effect, and requires a more elaborate mathematical and numerical
treatment to uncover its intrinsic complexity. In the limit $\Lambda \gg 1$,
the electron mean free path is large compared with the skin depth, and the
electron current is determined not by the local rf electric field (Ohm's
law), but rather is a function of the whole profile of the rf electric field
over distances of order $\lambda $. Therefore, a rather complicated nonlocal
conductivity operator has to be determined for the calculation of the rf
electric field penetration into the plasma.

In the case of a uniform plasma, the Vlasov and Maxwell equations can be
solved by applying a Fourier transform \cite{Pippard}. For a transverse
harmonic wave in one-dimensional geometry $E_{y}(x)e^{-i\omega t},$ a
spatial Fourier harmonic of the current $j_{yk}\exp (-ikx)$ simplifies to
become \cite{Kolobov review, me and Oleg}
\begin{equation}
j_{yk}=\frac{e^{2}n}{imkV_{T}}Z\left( \frac{\omega }{kV_{T}}\right) E_{yk}.
\label{jk}
\end{equation}%
Details of the solution are given in Appendix C. The electric field profile
is given by the inverse Fourier transform of Eq.(\ref{Maxwell 1D})
\begin{equation}
E_{y}(x)=\frac{2i\omega }{c^{2}}I\int_{-\infty }^{\infty }\frac{e^{ikx}}{%
k^{2}-\omega ^{2}\varepsilon _{t}(\omega ,k)/c^{2}}dk.  \label{Ey anomalous}
\end{equation}%
Here, $I$ is the surface current in the antenna and $\varepsilon _{t}(\omega
,k)$ is transverse plasma permittivity, which for a Maxwellian EEDF is given
by \cite{Lifshitz and Pitaevskii}
\begin{equation}
\varepsilon _{t}(\omega ,k)\simeq 1+\frac{\omega _{p}^{2}}{\omega ^{2}}\frac{%
\omega }{v_{T}|k|}Z\left( \frac{\omega }{v_{T}|k|}\right) .
\label{transverse permittivity}
\end{equation}%
Note the module sign as an argument of the plasma dispersion function. It
reflects the proper symmetry of the continued electric field profile into
semi-plane $x<0$ and also the proper pole position of the plasma dispersion
function \cite{Landau, Aliev and me}. Neglecting the module sign results in
erroneous results.

The solution for the electric field Eq.(\ref{Ey anomalous}) has been
described in many reviews and textbooks \cite{Lifshitz and Pitaevskii, Aliev
and me, Lieberman& Godyak review, Kolobov review}. Here, we only focus on a
property of the solution (\ref{Ey anomalous}) not commonly acknowledged in
the literature.

In the limit $\Lambda \gg 1$ or $\delta \ll v_{T}/\omega $, the plasma
dielectric function can be substituted by its limiting value at small
arguments $Z\simeq i\sqrt{\pi }$. Introducing the anomalous skin depth
\begin{equation}
\delta _{a}\equiv \frac{c}{\omega _{p}}\left( \frac{\omega _{p}v_{T}}{\omega
c\sqrt{\pi }}\right) ^{1/3},
\end{equation}%
and substituting $Z\simeq i\sqrt{\pi }$ into Eq.~(\ref{transverse
permittivity}) and into Eq.~(\ref{Ey anomalous}) gives
\begin{equation}
E_{y}(x)=\frac{2i\omega }{c^{2}}I\int_{-\infty }^{\infty }\frac{e^{ikx}}{%
k^{2}-i/|k|\delta _{a}^{3}}dk.  \label{Ey anomolous big limit}
\end{equation}

\begin{figure}[th]
\includegraphics[width=3in]{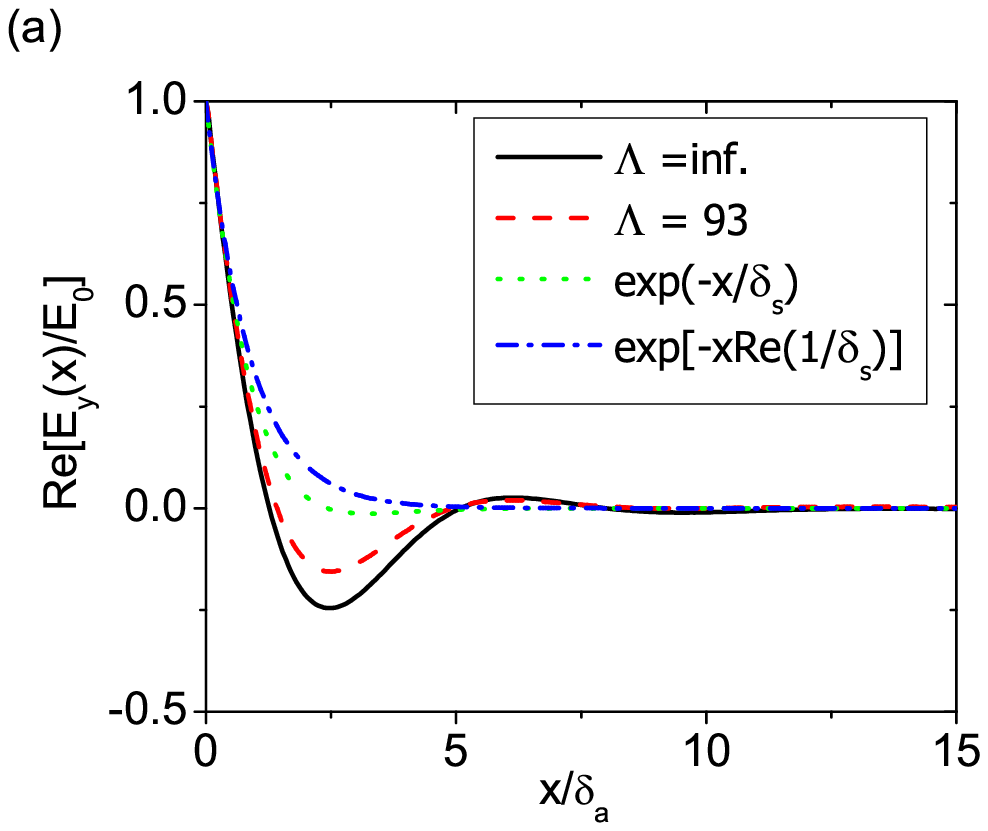},%
\includegraphics[width=3in]{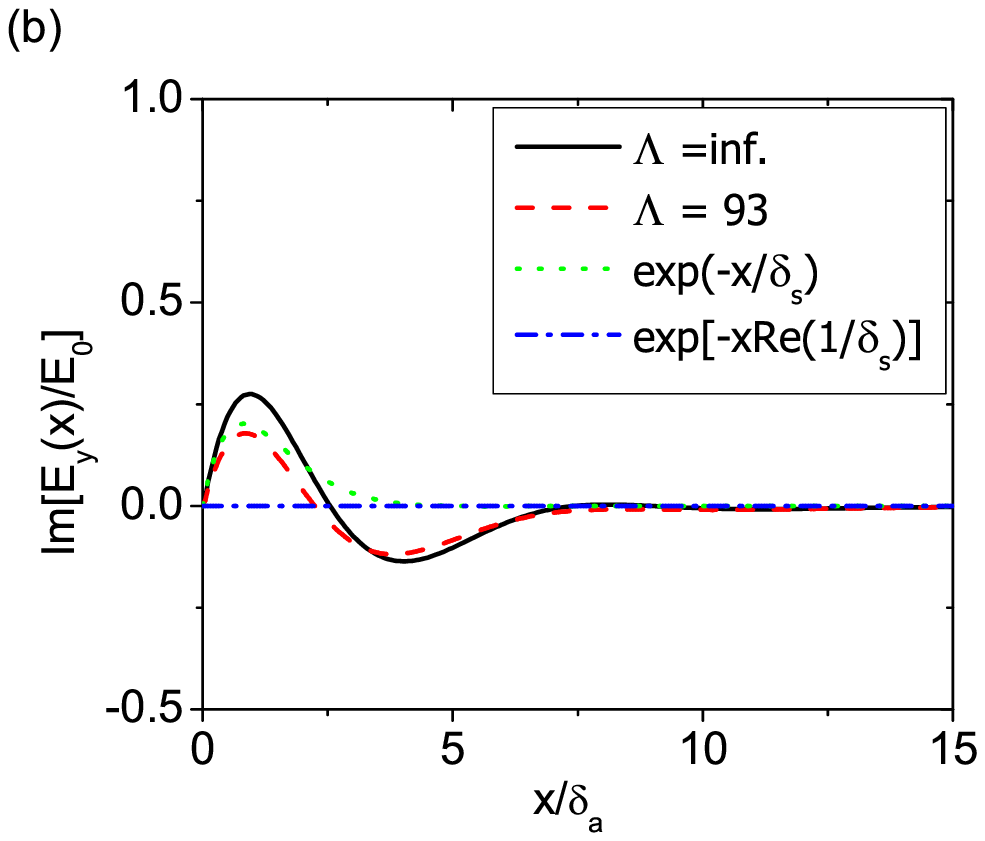}
\caption{Plot of the rf electric field as a function of the normalized
coordinate $x$/$\protect\delta _{a}$. The solid curve corresponds to the
solution in the limit $\Lambda =v_{T}\protect\omega _{p}/c\protect\omega %
=\infty $; dashed line - $\Lambda =93$ (plasma parameters n = 10$^{11}$cm$%
^{-3},$ T$_{e}$ = 3 eV, f = 1 MHz). The dotted and dash-dotted lines shows
the skin approximation in Eqs.~(\protect\ref{skin depth estimate}) and (%
\protect\ref{impedance approximation}): (a) real, and (b) imaginary part of
the electric field.}
\label{anomolous profile}
\end{figure}

\begin{figure}[th]
\includegraphics[width=3in]{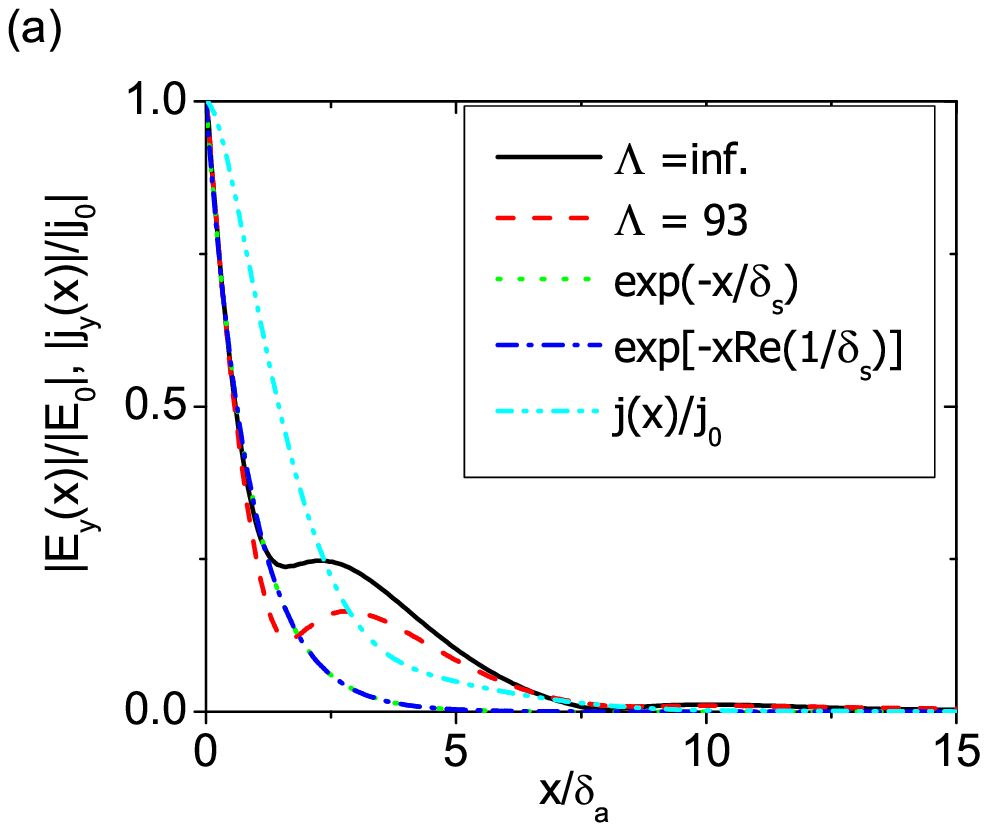},%
\includegraphics[width=3in]{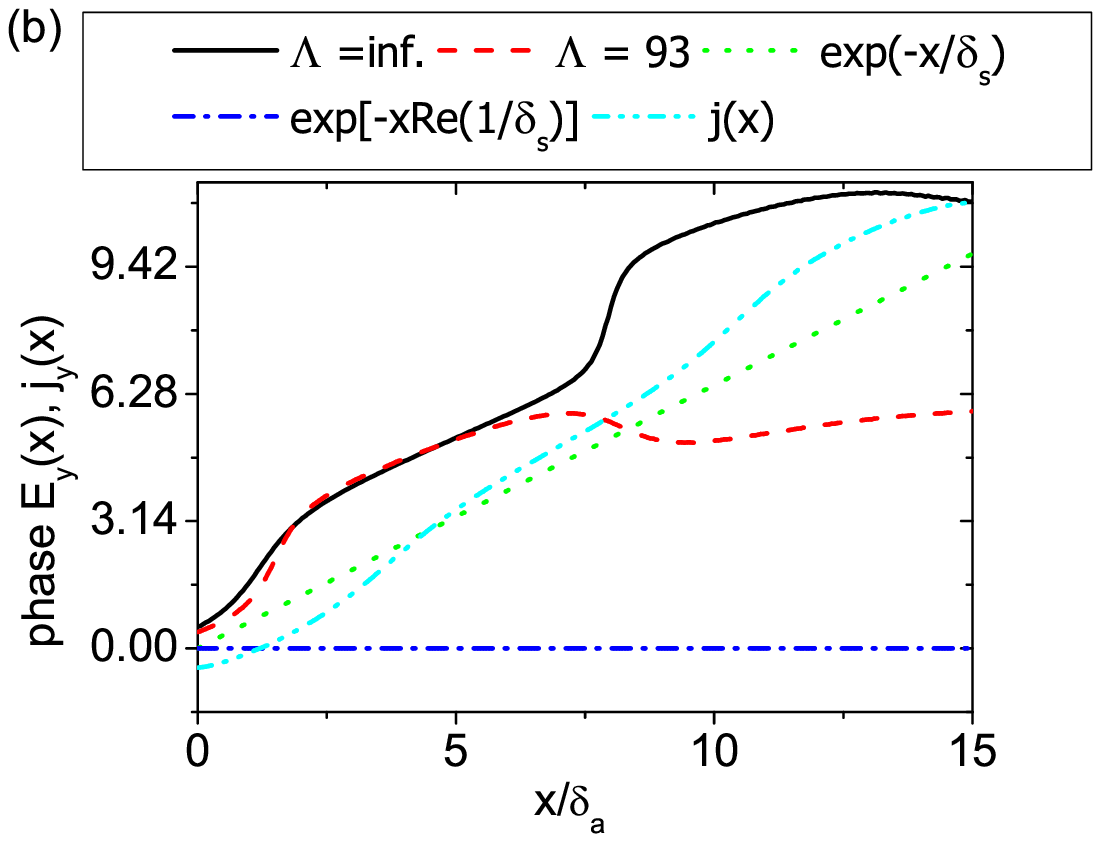}
\caption{Plot of the rf electric field and electron current as a function of
the normalized coordinate $x$/$\protect\delta _{a}$. The same profiles as in
Fig.~\protect\ref{anomolous profile}, shown are (a) amplitude, and (b) phase
with respect to the phase of the electric field generated by the field in
vacuum. }
\label{Eamplitude}
\end{figure}

The integral in Eq.~(\ref{Ey anomolous big limit}) cannot be calculated
analytically, but it can be transformed into an integral in the complex $k$
plane by substituting $|k|=\sqrt{k^{2}}$. The contour of the integration
should encompass branch point of the function $\sqrt{k^{2}}$ and has to come
around the imaginary $k-$axis. This gives \cite{Aliev and me}
\begin{eqnarray}
E_{y}(x) &=&E_{0}\frac{(i\sqrt{3}+1)}{3\gamma _{1}}\exp \left( -\frac{%
x\gamma _{2}}{\delta _{a}}\right) +\frac{E_{0}}{3\gamma _{1}}\exp \left( -%
\frac{x}{\delta _{a}}\right)  \label{anomolous skin profile} \\
&&-\frac{2iE_{0}}{\pi \gamma _{1}}P\int_{0}^{\infty }\frac{\xi \exp \left(
-x\xi /\delta _{a}\right) }{1-\xi ^{6}}d\xi .
\end{eqnarray}%
where $\gamma _{1}=2(\sqrt{3}+i)/3\sqrt{3},$ $\gamma _{2}=(1-i\sqrt{3})/2$
and $E_{0}$ is the electric field at the plasma boundary at $x=0$, $P$
stands for principal value of the integral. The last term represents the
contribution of the integral around the imaginary $k-$axis and the
exponential terms originate from the poles. The electric field at $x=0$ can
be calculated analytically
\begin{equation}
E_{0}=\frac{4i\omega I}{c^{2}}\frac{\pi (\sqrt{3}+i)\delta _{a}}{3^{3/2}}.
\label{Ex0}
\end{equation}%
From Maxwell's equations it follows that the magnetic field near the coil is
$B|_{0+}=2\pi I/c$. Correspondingly, the derivative of the electric field at
the plasma boundary is%
\begin{equation}
\frac{dE_{y}}{dx}|_{x=0}\mathbf{=-}\frac{2\pi i\omega }{c^{2}}I.
\end{equation}%
The characteristic decay length of the electric field can be introduced as
\cite{Kondratenko, Rukhadze}
\begin{equation}
\delta _{s}=\frac{E_{0}}{-dE_{y}/dx}\mathbf{=}\frac{2}{3}\left( 1+i/\sqrt{3}%
\right) \delta _{a}.  \label{skin depth estimate}
\end{equation}%
The electric field profile from Eq.~(\ref{anomolous skin profile}) is
compared in Fig.~\ref{anomolous profile} with the exponential profile
\begin{equation}
E_{y}(x)=E_{0}\exp \left[ -xRe(1/\delta _{s})\right] .
\label{impedance approximation}
\end{equation}%
A more conventional plot of the amplitude and phase of the electric fields
is shown in Fig.~\ref{Eamplitude}.

\subsection{Spatially averaged electric field, $\protect\int _{0}^{\infty
}E_{y}dx\rightarrow 0$ in the limit of a strong anomalous skin effect $%
\Lambda \rightarrow \infty $.}

The most apparent difference between the anomalous skin effect and the
normal skin effect is that the amplitude of the rf filed is non-monotonic in
the limit of anomalous skin effect and monotonic (exponential) for the
normal skin effect. Moreover, in the case of the extremely anomalous skin
effect, in the limit $\Lambda \gg 1$, the spatially averaged rf electric
field tends to zero \cite{Aliev and me}
\begin{equation}
\int_{0}^{\infty }E_{y}dx\rightarrow 0,\Lambda \rightarrow \infty .
\label{zero average}
\end{equation}%
In other words, the phase of the electric field changes by $\pi $ inside the
skin layer, see Fig.~\ref{Eamplitude}(b). The spatially averaged electric
field is given by the Fourier component at $k=0$, i.e.,
\begin{equation}
\int_{0}^{\infty }E_{y}dx=\pi E(k=0) .  \label{averaged E}
\end{equation}%
Substituting the Fourier component of the electric field from Eq.(\ref{Ey
anomalous}) into Eq.(\ref{averaged E}) gives
\begin{equation}
\int_{0}^{\infty }E_{y}(x)dx=\frac{2\pi i\omega I}{c^{2}}\frac{1}{\omega
^{2}\varepsilon (\omega )/c^{2}},
\end{equation}%
and
\begin{equation}
\frac{\int_{0}^{\infty }E_{y}(x)dx}{|E_{0}|\delta _{a}}=\frac{3^{3/2}\pi
^{1/3}}{\Lambda ^{1/3}}.  \label{integral
of E anmolous}
\end{equation}%
From Eq.~(\ref{integral of E anmolous}) it is evident that as the
nonlocality parameter tends to infinity, the averaged electric field tends
to zero. This property of the electric field profile is consistent with
nonlocality of the electron current. The electric field profile and the
current profile are coupled to each other by Eq.~(\ref{Maxwell 1D}).
Therefore, the main part of the current and the electric field should decay
on distances of order $\delta _{a}$, see Fig. \ref{Eamplitude}. However, if
the electric field profile has a non-zero average, the fast electrons will
pick up a velocity kick from the skin layer and will transport the current
over distances of order $v_{T}/\omega \gg \delta _{a}$, where the electric
field vanishes. This would contradict Maxwell's equations. \emph{Therefore,
the zero average of the electric field is necessary and an important
property of the electric field profile in the limit of the extreme anomalous
skin effect }$\Lambda \rightarrow \infty $.

The penetration length is defined in textbooks \cite{Kondratenko, Rukhadze}
as
\begin{equation}
\lambda _{E}=\frac{\int_{0}^{\infty }E_{y}(x)dx}{E_{0}}.
\end{equation}%
From the above discussion it follows that this definition is confusing,
because in the limit of the anomalous skin effect the above defined
penetration length is $\lambda _{E}\ll \delta _{a}$ and is not a good
measure of penetration length of the electric field. A better definition
would be
\begin{equation}
\lambda _{|E|}=\frac{\int_{0}^{\infty }|E_{y}(x)|dx}{|E_{0}|}.
\label{penetration length}
\end{equation}%
In the limit of the strong anomalous skin effect, i.e. $\Lambda \gg 1$,
numerical calculation gives
\begin{equation}
\lambda _{|E|}=1.64\delta _{a}.
\end{equation}%
From Fig.~\ref{Eamplitude} it is evident that in the region $x\lesssim
2\delta _{a}$ the amplitude of the electric field can be approximated by the
exponential profile in Eq.(\ref{impedance approximation}) with the decay
length
\begin{equation}
\delta _{e}=\frac{1}{Re(1/\delta _{s})}=\frac{8}{9}\delta _{a}.
\end{equation}%
Note that the penetration length defined by Eq.~(\ref{penetration length}), $%
\lambda _{|E|}$ is nearly twice as large as the initial decay length of the
electric field amplitude near the plasma-wall boundary $\delta _{e}$. This
is due to the pronounced long tail in the profile of the electric field.

Similarly, if we introduce the penetration length of the current
\begin{equation}
\lambda _{|j|}=\frac{\int_{0}^{\infty }|j_{y}(x)|dx}{|j_{0}|},
\end{equation}%
numerical simulation gives
\begin{equation}
\lambda _{|j|}=1.87\delta _{a}\approx \lambda _{|E|}.
\end{equation}%
This result contradicts to claim of Refs.\cite{Kondratenko,Godyak review},
that the magnetic field and current penetration lengths are much longer than
the electric field penetration length. This claim is the result of an
inaccurate definition of the penetration length.

In an attempt to reduce the phenomenon of the anomalous skin effect to the
normal skin effect, many authors have substituted the correct profile of the
electric field in Eq.~(\ref{anomolous skin profile}) by an exponential
profile $E_{0}\exp (-x/\delta _{e})$ with some fitting procedure for $\delta
_{e}$ \cite{Vahedi, Haas, Tyshetskiy}. By doing so, the property of the
electric field in the limit of anomalous skin effect in Eq.(\ref{zero
average}) is violated. This leads to overestimation of the electron heating
\cite{Aliev and me}. Under the conditions of the anomalous skin effect $%
v_{T}\gg \delta _{a}\omega $, electrons acquire a velocity kick
\begin{equation}
\Delta v_{y}=-\frac{2e}{mv_{x}}\int_{0}^{\infty }E_{y}(x)dx.
\end{equation}%
If $E_{y}(x)$ satisfies the condition in Eq.~(\ref{zero average}), the
electron velocity kick after passing through the skin layer is much smaller
than in the case of an exponential electric field profile, which does not
satisfy the property $\int_{0}^{\infty }E_{y}dx\rightarrow 0$, as $\Lambda
\rightarrow \infty $.

\subsection{Analytical separation of the electric field profile into an
exponential part and a far tail.}

Consider an exponential profile of the electric field in a plasma
\begin{equation}
E_{y}(x,t)=E_{y0}\exp \left( -k_{p}x-i\omega t\right) ,
\label{exponentional profile}
\end{equation}%
where $k_{p}$ is a real positive number. The velocity perturbation in this
electric field becomes
\begin{equation}
\Delta v_{y}(x,t)=-\frac{e}{m}\int_{-\infty }^{t}d\tau E_{y}[x(\tau ),\tau ].
\end{equation}%
The velocity kick $\Delta v_{y}$ can be separated into a purely exponential
part and a non-exponential part. Substituting the electron trajectory $%
x(\tau )=x-v_{x}(t-\tau )$ for $v_{x}<0$ gives
\begin{equation}
\Delta v_{y}(x,t)=-\frac{e}{m}\frac{Ey_{0}}{-k_{p}v_{x}-i\omega }\exp \left(
-k_{p}x-i\omega t\right) .  \label{v kick}
\end{equation}%
For $v_{x}>0$, the velocity acquired by an electron can be represented as
the difference between the velocity kick acquired after a full pass through
the skin layer and the contribution from the part of the skin layer $%
[x;\infty ]$, i.e.,
\begin{equation}
\Delta v_{y}(x,t)=-\frac{e}{m}\left[ \int_{-\infty }^{\infty
}-\int_{t}^{\infty }\right] d\tau E_{y}[x(\tau ),\tau ].  \label{dvy vm}
\end{equation}%
The second part of the integral ($\Delta v_{y}^{e}$) in Eq.~(\ref{dvy vm})
gives an exponential profile for the velocity kick, similar to Eq.~(\ref{v
kick})
\begin{equation}
\Delta v_{y}^{e}(x,t)=-\frac{e}{m}\frac{E_{y0}}{-k_{p}v_{x}-i\omega }\exp
\left( -k_{p}x-i\omega t\right) ,\;v_{x}>0.  \label{v kick 2}
\end{equation}%
The first part of the integral ($\Delta v_{y}^{in}$) in Eq.~(\ref{dvy vm})
gives
\begin{equation}
\Delta v_{y}^{in}=\Delta v_{y}^{\infty }e^{-i\omega (t-x/v_{x})},
\label{v_kick_nonexp}
\end{equation}%
\begin{equation}
\Delta v_{y}^{\infty }=-\frac{e}{m}E_{y0}\left( \frac{1}{-i\omega +k_{p}v_{x}%
}-\frac{1}{-i\omega -k_{p}v_{x}}\right) .  \label{Dvy}
\end{equation}%
Here, $\Delta v_{y}^{\infty }$ is the velocity kick acquired during the pass
through the entire skin layer. The time $t-x/v_{x}$ corresponds to the
moment the electron collides with the wall.

Substituting $\Delta v_{y}^{e}(x,t)$ from Eqs.~(\ref{v kick}) and (\ref{v
kick 2} ) gives for the exponential part of the current%
\begin{equation}
j_{y}^{e}=-e\int \Delta v_{y}\frac{\partial f}{\partial v_{y}}v_{y}d\mathbf{%
v,}
\end{equation}%
\begin{equation}
j_{y}^{e}=\frac{e^{2}}{m}E_{y0}e^{-i\omega t-k_{p}x}\int \frac{1}{%
-k_{p}v_{x}-i\omega }\frac{\partial f}{\partial v_{y}}v_{y}d\mathbf{v,}
\end{equation}%
or, after integration, the exponential profile of the current becomes%
\begin{equation}
j_{y}^{e}=\frac{e^{2}}{m}E_{y0}e^{-i\omega t-k_{p}x}\frac{n}{k_{p}V_{T}}%
Z\left( \frac{i\omega }{k_{p}V_{T}}\right) ^{\ast }.  \label{jc k}
\end{equation}%
The asterisk denotes the complex conjugate. Note that Eq.~(\ref{jc k}) can
be derived from Eq.~(\ref{jk}) with the substitution $k=ik_{p}$ and by
accounting for the following property of the dispersion function \cite%
{Plasma Formulary}
\begin{equation}
Z(\xi ^{\ast })=-Z(-\xi )^{\ast }.
\end{equation}%
The exponential part of the profile should satisfy Maxwell's equation (\ref%
{Maxwell 1D}). This gives an expression for $k_{p}$
\begin{equation}
k_{p}^{2}\mathbf{=}\frac{\omega ^{2}}{c^{2}}+\frac{\omega _{p}^{2}}{c^{2}}%
\frac{i\omega }{k_{p}V_{T}}Z\left( \frac{i\omega }{k_{p}V_{T}}\right) ^{\ast
}.  \label{kp}
\end{equation}%
Note that because $Z$ in Eq.~(\ref{kp}) has only purely imaginary and
positive parts, $k_{p}$ is a real positive number, as it was assumed to be.

The non-exponential part of the electron velocity kick in Eq.(\ref%
{v_kick_nonexp}) generates a non-exponential part of the current profile,
which decays over a spatial scale of order $V_{T}/\omega $ due to the phase
mixing, as the phase of the velocity kick $\omega (t-x/v_{x})$ in Eq.~(\ref%
{v_kick_nonexp}) is different for electrons with different $v_{x}.$ The
current and electric field profiles are essentially non-exponential, similar
to Eq.(\ref{current approx}) for longitudinal velocity kicks, as discussed
above.

Details of the exact analytical calculation of the electric field profile
separation is given in Appendix C. Applying a procedure similar to that of
Landau's treatment\cite{Landau} for the longitudinal electric field, the
integral in $k-$space in Eq.~(\ref{Ey anomalous}) can be separated into an
integral over an analytic function in the region $k\in \lbrack -\infty
,\infty ]$ and an integral over some non-analytic function in the region $%
k\in \lbrack 0,\infty ]$. To do so, the plasma permittivity has to be
analytically continued from the real axis $k<0$, $Imk=0$, into the complex $%
k $ -plane, see Appendix C for details. The first integral can be readily
calculated using the theory of residues. In the upper half-plane of the
complex $k$, there exists only one pole of the analytically continued
function of the plasma permittivity continued from $k<0$. The value of the
pole is equal to $ik_{p}$, given by Eq.~(\ref{kp}).

In the limit $\omega \gg k_{p}V_{T}$, $Z(\zeta )=-1/\zeta $, where $\zeta
=i\omega /k_{p}V_{T}$. Substituting this value for the plasma dielectric
function into Eq.~(\ref{kp}) yields $k_{p}=\omega _{p}/c$, i.e., the normal
skin layer length $1/k_{p}=\delta _{0}$ in Eq.~(\ref{normal skin effect})
for $\nu \ll \omega $ and $\omega \ll \omega _{p}.$
\begin{figure}[th]
\includegraphics[width=3in]{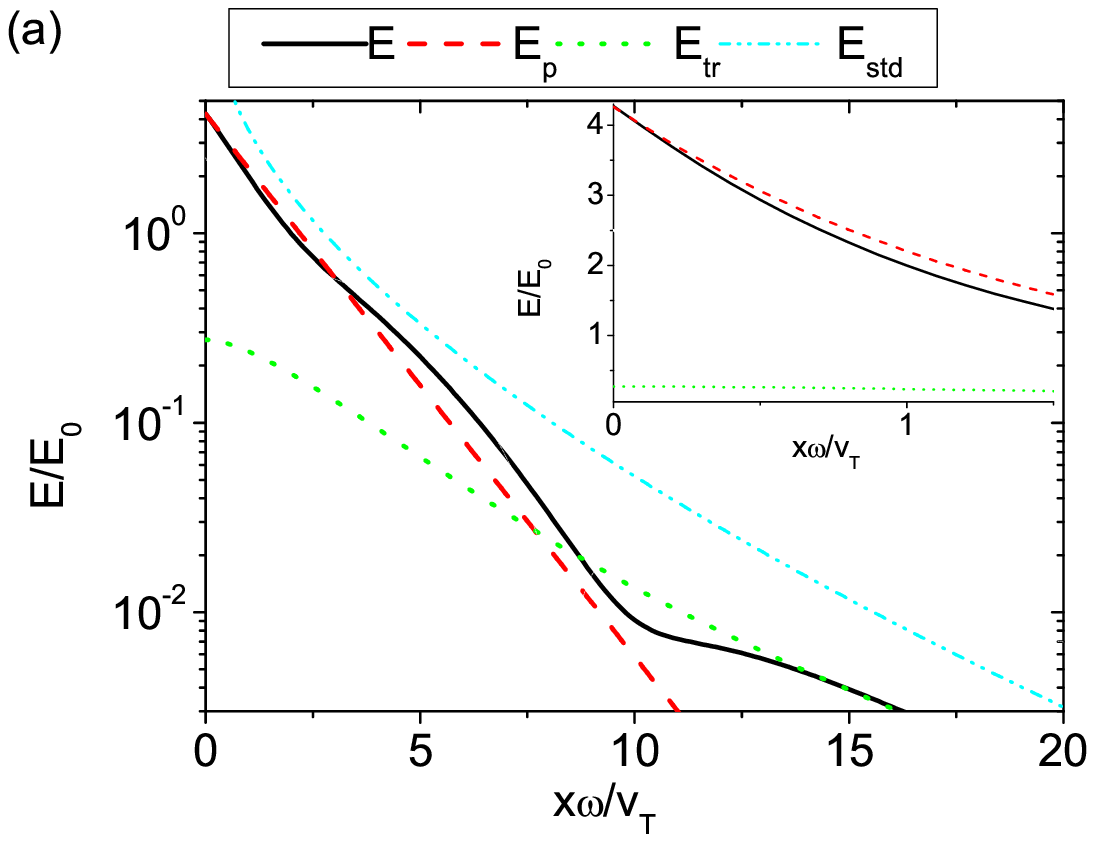}, %
\includegraphics[width=3in]{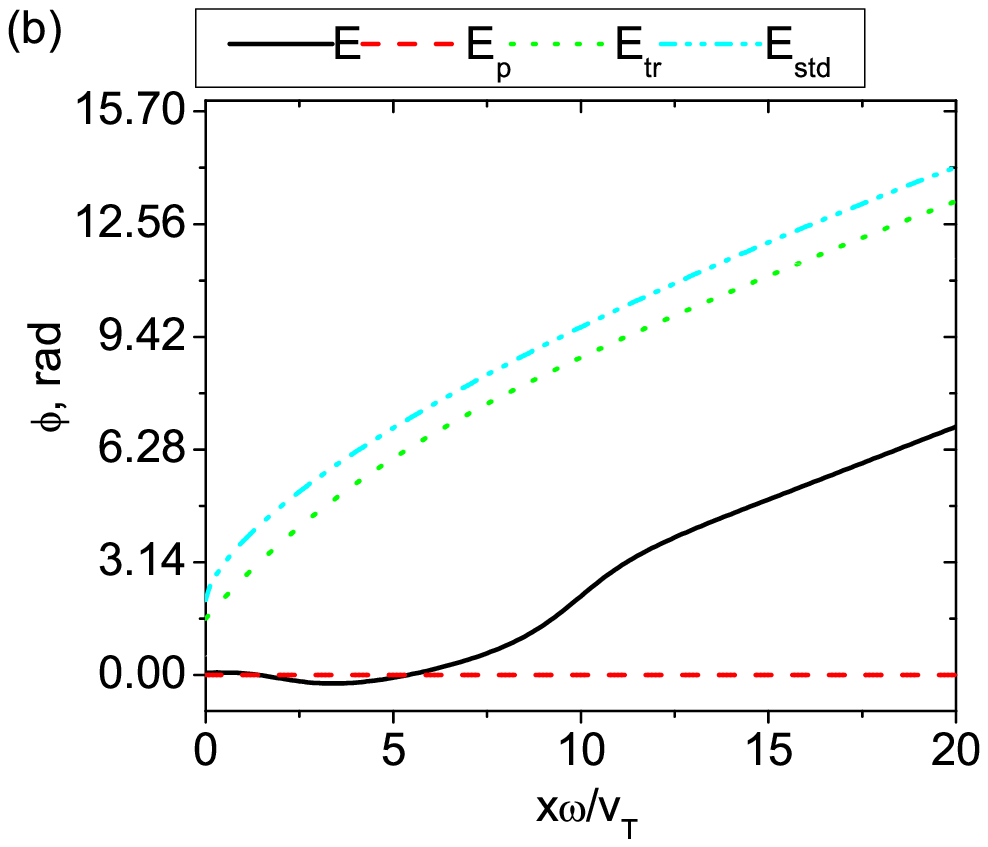}
\caption{Plot of the rf electric field as a function of the normalized
coordinate $x\protect\omega $/$V_{T}$ for plasma parameters $n=10^{11}$cm$%
^{-3},T_{e}=3$eV, $f=13.56$ MHz. Shown are (a) the amplitude and (b) the
phase. Solid lines show the exact electric field profile $E(x)$ calculated
according to Eq.~(\protect\ref{Ey anomalous}); dashed (red) line, the
exponential part of the electric field $E_{p}(x)=E_{0}\exp (-k_{p}x)$ with $%
k_{p}$ from Eq.~(\protect\ref{kp}); dotted line (green), the difference of
the two $E_{t}(x)$; and, chain (cyan) line, $E_{std}(x)$ shows the
asymptotic calculation for $E_{t}$ in Eq.~(\protect\ref{std anomolous skin}%
). Subscript $y$ is ommitted in the electric field.}
\label{Figure f13}
\end{figure}
Figure \ref{Figure f13} shows the profile of the electric field for the same
typical ICP parameters: plasma density $n=10^{11}$cm$^{-3},$ electron
temperature $T_{e}=3$ eV, and discharge frequency $f=13.56$ MHz. Shown are
the exact electric field profile $E_{y}(x)$ calculated according to Eq.(\ref%
{Ey anomalous}), the exponential part of the electric field
\begin{equation}
E_{yp}(x)=E_{0}\exp (-k_{p}x)  \label{Ep}
\end{equation}%
with $k_{p}$ from Eq.(\ref{kp}), and the difference of the two
\begin{equation}
E_{yt}(x)=E_{y}(x)-E_{yp}(x),
\end{equation}%
and the asymptotic calculation for $E_{yt}(x)$ in Eq.(\ref{std anomolous
skin} ) $E_{ystd}(x)$. For these plasma parameters the skin effect is
neither normal nor anomalous: $\omega /k_{p}V_{T}=1.52$. Notwithstanding the
fact that the parameter $\omega /k_{p}V_{T}$ is of order unity, the main
part of the electric field is close to the exponential profile in Eq.~(\ref%
{Ep}) with $k_{p}$ from Eq.~(\ref{kp}), $E_{y}(x)\approx E_{yp}(x)$. As
evident from Fig.~\ref{Figure f13}, the non-exponential part is small,$%
E_{yt}(x)\ll E_{yp}(x)$, everywhere where the electric field is substantial,
or up to distances five times of skin depth, for $x<5/k_{p}=7.5V_{T}/\omega $%
. The tail of the electric field profile for $x>7V_{T}/\omega $ is
non-exponential and dominated by $E_{yt}(x)$.

\begin{figure}[th]
\includegraphics[width=3in]{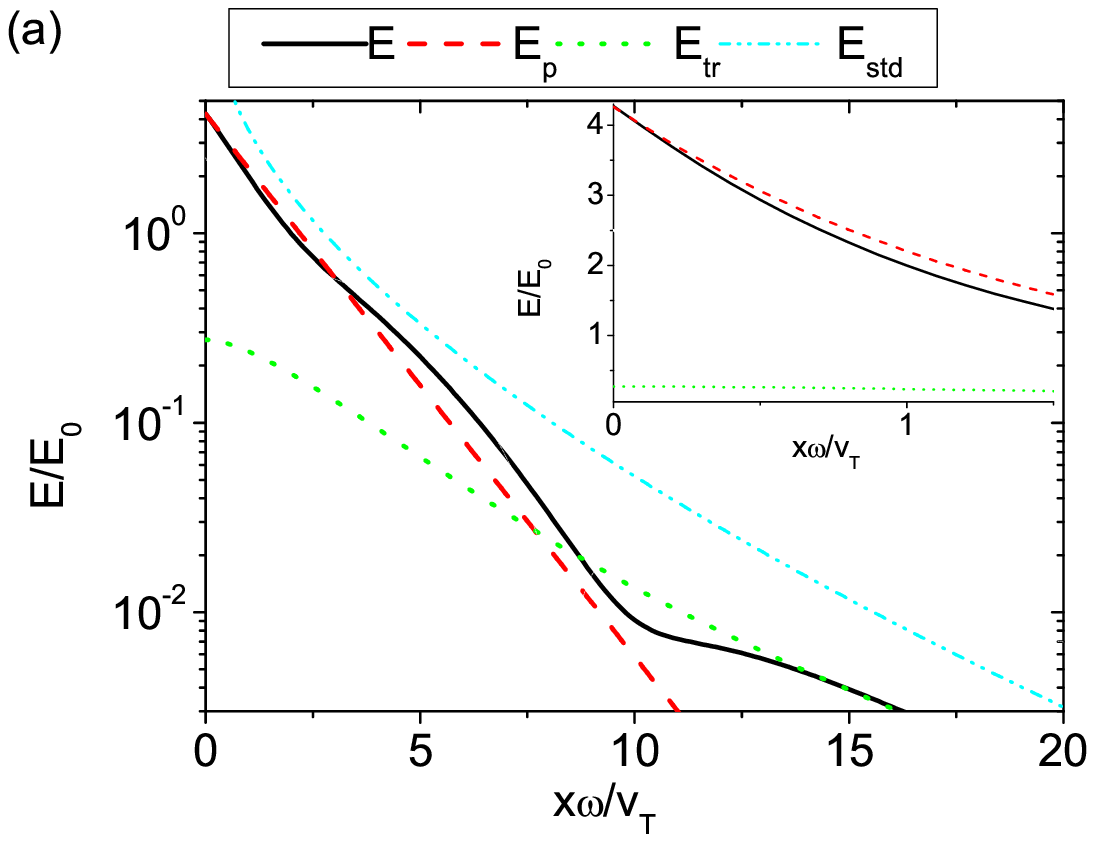}, %
\includegraphics[width=3in]{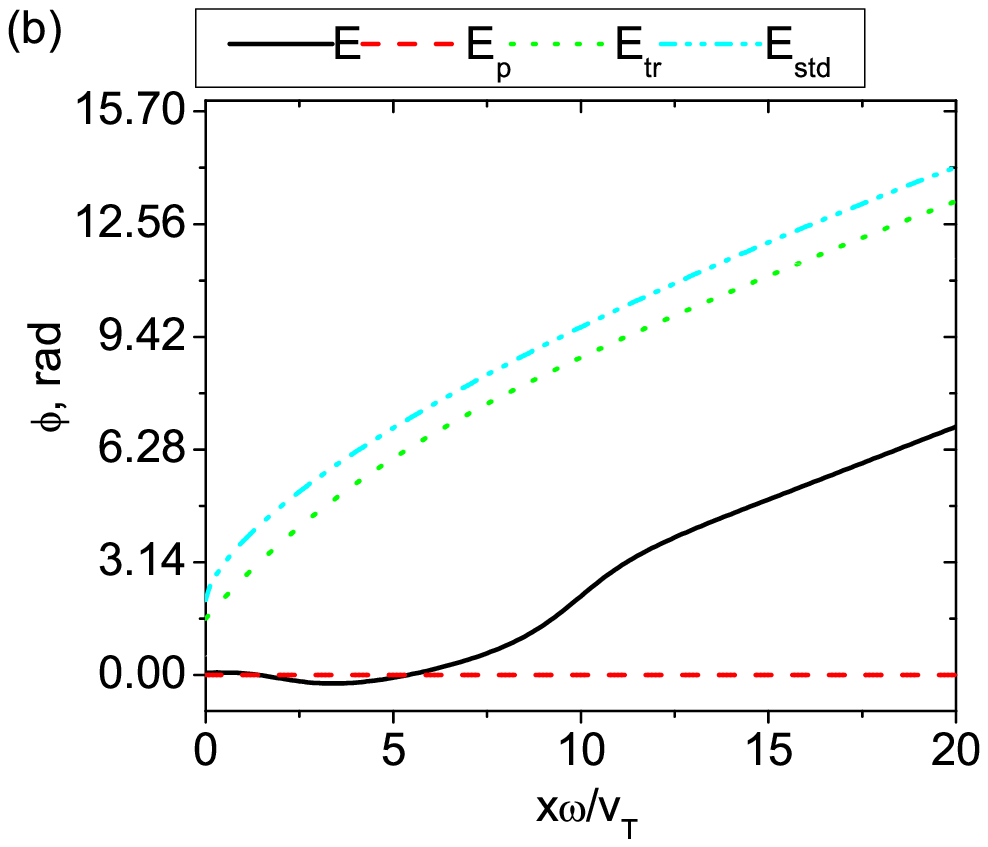}
\caption{Plot of the rf electric field as a function of the normalized
coordinate $x\protect\omega $/$V_{T}$ for plasma parameters $n=10^{11}$cm$%
^{-3},T_{e}=3$ eV, $f=1$ MHz, similar to Figs.~\protect\ref{anomolous
profile} and \protect\ref{Eamplitude}. Shown are (a) amplitude and (b)
phase. Solid lines show the exact electric field profile $E(x)$ calculated
according to Eq.~(\protect\ref{Ey anomalous}); dashed (red) line, the
exponential part of the electric field $E_{p}(x)=E_{0}\exp (-k_{p}x)$ with $%
k_{p}$ from Eq.~(\protect\ref{kp}); dotted line (green) the difference of
the two $E_{t}(x)$; chain (blue) line represents the limiting case of strong
anomalous skin effect $\Lambda \rightarrow \infty $ $E_{appr}(x)$, and
dashed and double dotted (chain) line shows $E_{std}(x)$, the asymptotic
calculation for $E_{t}$ in Eq.~(\protect\ref{std anomolous skin}). Subscript
$y$ for the electric fields is omitted.}
\label{Figure f1}
\end{figure}

In the limit of the anomalous skin effect $\omega /k_{p}V_{T}\ll 1$, $%
Z(\zeta )=i\sqrt{\pi }$, where $\zeta =i\omega /k_{p}V_{T}$. Substituting
this value for the plasma dielectric function into Eq.~(\ref{kp}) yields $%
k_{p}=1/\delta _{a},$ which is very close to the skin impedance
approximation in Eq.~(\ref{impedance approximation}) which corresponds to $%
k_{p}=9/8\delta _{a}$ --a 12 \% difference. As a result, the exponential
profile in Eq.~(\ref{Ep}) approximates well the exact profile of the
electric field over distances within a few skin depths even in the limit of
the strong anomalous skin effect, as is evident in Fig.~\ref{Figure f1}.
However, the non-exponential part $E_{yt}(x)$ dominates $E_{yp}(x)$ at $%
x>V_{T}/\omega $ in accord with the requirement in Eq.(\ref{zero average}).

\subsection{Surface impedance}

An important plasma characteristic is the surface impedance, which is given
by the ratio of the electric field to the rf magnetic field or the coil
current at the plasma boundary \cite{Lifshitz and Pitaevskii}
\begin{equation}
Z =\frac{E}{B}|_{x=0},  \label{surface impedance}
\end{equation}%
where
\begin{equation}
B|_{x=0}=\frac{2\pi }{c}I  \label{Eq. B}
\end{equation}%
is the magnetic field near the antenna. The total power $P$ deposited per
unit area into the plasma is determined by the energy flux dissipated into
the plasma or the time-averaged Poynting vector
\begin{equation}
P=<S_{x}>=\frac{1}{2}\frac{c}{4\pi }Re(EB^{\ast }).  \label{Poynting}
\end{equation}%
Substituting the electric field from Eq.(\ref{surface impedance}) and the
magnetic field Eq.~(\ref{Eq. B}) into Eq.~(\ref{Poynting}) relates the power
to the real part of the surface impedance
\begin{equation}
P=\frac{\pi }{2c}I^{2}ReZ.  \label{power from impedance}
\end{equation}%
The imaginary part of the surface impedance describes the plasma inductance.

The surface impedance can also be used to estimate the penetration length in
the surface impedance approximation given by Eq.~(\ref{skin depth estimate}%
). Substituting the electric field from Eq.~(\ref{surface impedance}) and
the magnetic field Eq.~(\ref{Eq. B}) into Eq.(\ref{skin depth estimate})
relates the penetration depth and the surface impedance%
\begin{equation}
\delta _{s}=\frac{cZ}{i\omega }.  \label{delta from Z}
\end{equation}%
The surface impedance can be calculated making use of Eq.~(\ref{Ey anomalous}%
) \cite{Lifshitz and Pitaevskii}, i.e.,
\begin{equation}
Z=\frac{i\omega }{\pi c}\int_{-\infty }^{\infty }\frac{1}{k^{2}-\omega
^{2}\varepsilon _{t}(\omega ,k)/c^{2}}dk, \label{impedance}
\end{equation}%
which requires numerical integration. On the other hand, we can use the
results of the previous subsection that the main part of the electric field
is an exponential function in Eq.~(\ref{Ep}) with $k_{p}$ given by Eq.(\ref%
{kp}). From Eq.~(\ref{delta from Z}), the imaginary part of the surface
impedance can be obtained substituting $\delta _{s}=1/k_{p}$
\begin{equation}
Z_{p}=\frac{i\omega }{ck_{p}}.  \label{impedance p}
\end{equation}%
A pure exponential profile yields only the imaginary part of the surface
impedance. The real part of the impedance can be calculated by computing the
power dissipated by electrons from the skin layer \cite{Vahedi}
\begin{equation}
P=\frac{m}{4}\int v_{x}f|\Delta v_{y}^{\infty }|^{2}\mathbf{dv},
\label{power Ey}
\end{equation}%
where $\Delta v_{y}^{\infty }$ is the velocity kick acquired by an electron
after passing through the skin layer, which is given by Eq.~(\ref{Dvy}).
Here, $m(\Delta v_{y}^{\infty })^{2}/4$ is the temporal average of the
electron energy change in the skin layer and $v_{x}f$ is the electron flux
on the wall. Equation (\ref{power from impedance}) becomes
\begin{equation}
ReZ_{p}=\frac{2}{c}\omega _{p}^{2}|Z|^{2}\int fv_{x}\left( \frac{k_{p}v_{x}}{%
\omega ^{2}+(k_{p}v_{x})^{2}}\right) ^{2}dv_{x}.  \label{Real of Z}
\end{equation}%
Because the imaginary part of impedance is large compared with its real
part, only the imaginary part can be included on the right hand side in Eq.~(%
\ref{Real of Z}).
\begin{figure}[th]
\includegraphics[width=3in]{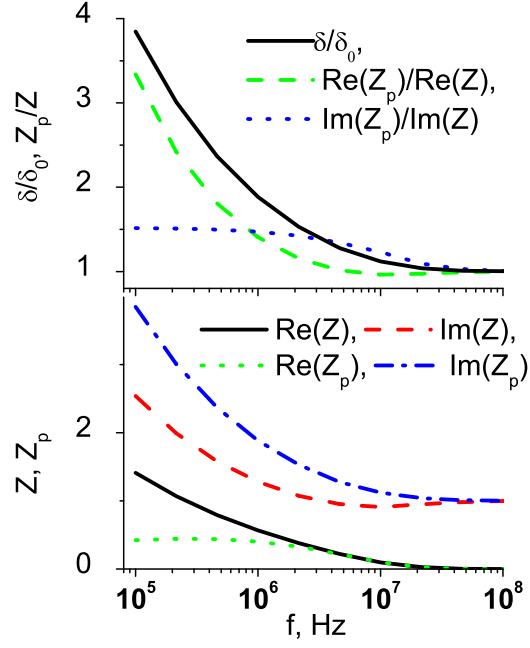}
\caption{Plot of the real and imaginary parts of the surface impedance
versus discharge frequency calculated exactly making use of Eq.~(\protect\ref%
{impedance}) and approximately using Eqs.~(\protect\ref{impedance p}) and (%
\protect\ref{Real of Z}) in the limit of collisionless plasma $\protect\nu %
\ll \protect\omega $. Also shown is the ratio of the actual skin depth $%
\protect\delta =1/k_{p}$ given by Eq.~(\protect\ref{kp}) to the skin depth
calculated in the cold plasma approximation $\protect\delta _{0}$ Eq.~(%
\protect\ref{normal skin effect}), (top). }
\label{Figure impedance on f}
\end{figure}
Figure \ref{Figure impedance on f} shows the real and imaginary parts of the
surface impedance versus the discharge frequency calculated exactly, i.e.,
making use of Eq.(\ref{impedance}), and approximately from Eqs.~(\ref%
{impedance p}) and (\ref{Real of Z}). Also shown at the top of this figure,
is the ratio of the actual skin depth $\delta =1/k_{p}$ from Eq.~(\ref{kp})
to the normal skin depth calculated in the cold plasma approximation $\delta
_{0}$ given by Eq.(\ref{normal skin effect}). From Fig.~\ref{Figure
impedance on f} it is evident that within 50\% accuracy, the impedance
calculation can be based on the exponential profile in Eq.~(\ref{Ep}) for
discharge frequencies higher than 1 MHz \cite{Haas}. However, for lower
frequencies, the assumption of purely exponential profile leads to
overestimation of the electron heating and plasma resistivity up to a factor
of 3 for $f\ll 1$ MHz, see Fig.~\ref{Figure impedance on f}. This is because
the important property of the electric field profile under the conditions of
strong anomalous skin effect in Eq.~(\ref{zero average}) is being violated.
Note that at these low frequencies taking into account a small but finite
collision frequency or nonlinear effects may be important.

\subsection{Anomalous skin effect for an anisotropic electron velocity
distribution}

\begin{figure}[tbp]
\includegraphics[width=70mm]{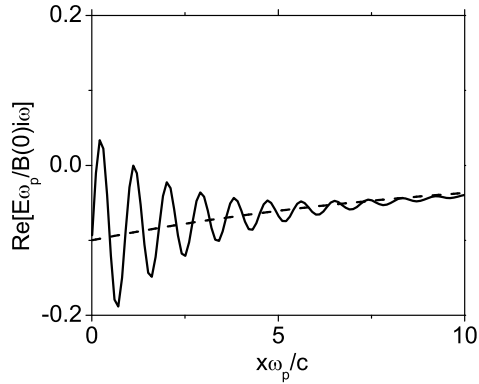}
\caption{{}The electric field in the plasma with $v_{Ty}=0.1c,$ $\protect%
\omega =0.01\protect\omega _{p}$, $T_{y}/T_{x}=50$. The solid line shows the
real part of the electric field profile obtained from the full solution. The
dashed line corresponds to the smooth part of the solution $\sim \exp (-%
\protect\omega x/v_{Ty})$. }
\label{Fig anisotropic skin}
\end{figure}

The anomalous skin effect in a plasma with a highly anisotropic electron
velocity distribution function (EVDF) is very different from the skin effect
in a plasma with the isotropic EVDF. In Ref. \cite{anisotropic EEDF} an
analytical solution was obtained for the electric field penetrating into
plasma with the EVDF described by a Maxwellian with two temperatures $%
T_{y}\gg T_{x}$, where $y$ is the direction along the plasma boundary and $x$
is the direction perpendicular to the plasma boundary. Under the conditions
\begin{equation}
\frac{v_{Ty}}{\omega }\gg \frac{c}{\omega _{p}};\;\omega _{p}\gg \omega ,
\label{Conditions1}
\end{equation}%
the skin layer was found to consist of two distinct regions of width of
order $v_{Tx}/\omega $ and $v_{Ty}/\omega $, where $v_{Tx,y}=\sqrt{T_{x,y}/m}
$ are the thermal electron velocities in $x$ and $y$ directions, and $\omega
$ is the incident wave frequency. The calculation is based on Eq.(\ref{Ey
anomalous}), where the dielectric permittivity has to be modified for an
anisotropic EEDF to become
\begin{equation}
\varepsilon _{t}(\omega ,k)=1-\frac{\omega _{p}^{2}}{\omega ^{2}}\left\{ 1-%
\frac{T_{y}}{T_{x}}\left[ 1+\frac{\omega }{\sqrt{2}v_{Tx}k}Z\left( \frac{%
\omega }{\sqrt{2}v_{Tx}k}\right) \right] \right\} .
\label{permitivity anisotropic}
\end{equation}%
In the case of anisotropic EEDF under conditions in Eq.~(\ref{Conditions1}),
the integral in Eq.(\ref{Ey anomalous}) has two poles and the integration
over the branch point $k=0$ does not contribute. As a result, the profile of
the electric field is a sum of the two complex exponents:
\begin{equation}
E(x)\simeq \frac{\omega }{\omega _{p}}B(0)\left[ -\frac{i\omega c}{\omega
_{p}v_{Ty}}\exp (ik_{p1}x)+\sqrt{T_{x}/T_{y}}\exp (ik_{p_{2}}x)\right] ,
\label{Eps}
\end{equation}%
where $k_{p1}$ is given by
\begin{equation}
k_{p1}=i\frac{\omega }{v_{Ty}},  \label{kp1}
\end{equation}%
and $k_{p2}$ is given by
\begin{equation}
k_{p2}=\frac{\omega _{p}}{c}\sqrt{T_{y}/T_{x}}+i\frac{\sqrt{\pi }\omega }{2%
\sqrt{2}v_{Tx}}.  \label{Imkp2}
\end{equation}%
The profile of the electric field is shown in Fig.~\ref{Fig anisotropic skin}%
. The skin layer contains multiple oscillations of the electric field, in
striking contrast to the case of isotropic EEDF.

\section{Conclusions}

We showed that electrons can transport the plasma current away from the skin
layer due to their thermal motion over distances of order $v_{T}/\omega $.
As a result, the width of the skin layer increases when electron temperature
effects are taken into account. The anomalous penetration of the rf electric
field occurs not only for the wave transversely propagating to the plasma
boundary (inductively coupled plasmas), but also for the wave propagating
along the plasma boundary (capacitively coupled plasmas). It was shown that
separating the electric field profile into exponential and nonexponential
parts yields an efficient qualitative and quantitative description of the
anomalous skin effect. Accounting for the non-exponential part of the
profile is important for the calculation of the electron heating and the
plasma resistivity. For example, the assumption of purely exponential
profile leads to overestimation of up to a factor of 3 in the electron
heating for $f\ll 1$ MHz, see Fig.~\ref{Figure impedance on f}.

Here, we considered only plasmas with a Maxwellian electron energy
distribution function. However, in low pressure rf discharges, the EEDF is
non-Maxwellian for plasma densities typically lower than $10^{10}$cm$^{-3}$
\cite{Godyak new exp}. The nonlocal conductivity, and plasma density
profiles and EEDF are all nonlinear and nonlocally coupled \cite{new Oleg}.
Hence, for accurate calculation of the discharge characteristics at low
pressures, the EEDF needs to be computed self-consistently \cite{badri and
me, badri and me 2, Our article, our1}. The effects of a nonMaxwellian EEDF,
nonlinear phenomena, the effects of plasma non-uniformity and finite size,as
well as influence of the external magnetic field on the anomalous skin
effect will be reported in the second part of the review \cite{review part 2}%
.

\textbf{Acknowledgments}

This research was supported by the U.S. Department of Energy Office of
Fusion Energy Sciences through a University Research Support Program and the
University of Toledo. The authors gratefully acknowledge helpful discussions
with R. Davidson, V. Godyak, Badri Ramamurthi, E. Startsev and L. D. Tsendin.

\appendix

\section{Analytical derivation of the current profile driven by velocity
kicks near the plasma boundary}

Consider that electrons acquire a velocity kick near the boundary, in the
direction perpendicular to the boundary
\begin{equation}
dv_{x}=\Delta V\cos (\omega t).
\end{equation}%
The electron velocity at a distance $x$ from the boundary will be determined
by the exact moment of the collision with the boundary at a time $t-x/v_{x}$%
. The electron current in the plasma is given by integration over all
electrons with a distribution function $f(v_{x})$
\begin{equation}
j(x,t)=e\Delta V\int_{0}^{\infty }f(v_{x})\cos (\omega t-\omega
x/v_{x})dv_{x}.  \label{current append}
\end{equation}%
For a Maxwellian distribution function $%
f(v_{x})=n_{0}e^{-v_{x}^{2}/v_{T}}/v_{T}\sqrt{\pi }$ the current in Eq.~(\ref%
{current append}) takes the form $j(\xi ,t)=j_{0}A(\xi )\cos [\omega t-\phi
(\xi )]$ , where $j_{0}=en_{0}\Delta V$ and $A$ and $\phi $ are the
amplitude and phase of the current, respectively, and $\xi =\omega x/v_{T}$.
The functions $A$ and $\phi $ are shown in Fig.~1. In the limit $\xi \gg 1$,
the integration in Eq.~(\ref{current append}) can be performed analytically
making use of the method of steepest descend \cite{Brillouin}
\begin{equation}
j(\xi ,t)=\frac{j_{0}}{\sqrt{\pi }}Re\left( e^{-i\omega t}\int_{0}^{\infty
}e^{-s^{2}+i\xi /s}ds\right) ,  \label{current integral append}
\end{equation}%
where and $s=v_{x}/v_{T}$. The integral in Eq.(\ref{current integral append}%
) can be calculated in the complex $s$ plane. The stationary phase point is
given by $d(-s^{2}+i\xi /s)/ds=0$ or $s^{3}=-i\xi /2$. This gives the
stationary point $s_{0}=\left( -i\xi /2\right) ^{1/3}$. In the neighborhood
of this point, the function in the exponent can be expanded as a Taylor
series, $-s^{2}+i\xi /s=-s_{0}^{2}+i\xi
/s_{0}-6(s-s_{0})^{2}/2=-3s_{0}^{2}-3(s-s_{0})^{2}$. Integration of the
Gaussian gives $\int_{0}^{\infty }e^{-3(s-s_{0})^{2}}ds=\sqrt{\pi /3}$.
Substituting this into the integration in Eq. (\ref{current integral append}%
) yields:
\begin{equation}
j(\xi ,t)=\frac{j_{0}}{\sqrt{3}}Re\left( \exp \left[ -i\omega t-3\left(
-i\xi /2\right) ^{2/3}\right] \right) .  \label{current integral 2}
\end{equation}%
Substituting $\left( -i\right) ^{2/3}=\left( e^{-i\pi /2}\right)
^{2/3}=e^{-i\pi /3}=\cos (\pi /3)-i\sin (\pi /3)=1/2-\sqrt{3}i/2$ into Eq.~(%
\ref{current integral 2}) gives
\begin{equation}
j(\xi ,t)=\frac{j_{0}}{\sqrt{3}}\exp \left( -3\xi ^{2/3}/2^{5/3}\right) \cos
\left( \omega t-3\sqrt{3}\xi ^{2/3}/2^{5/3}\right) .
\end{equation}

\section{Analytical derivation of the longitudinal rf electric field profile
near the plasma boundary ($\mathbf{E}\parallel \mathbf{k}$)}

The analytical solution for a longitudinal rf electric field involves
solving the Vlasov equation for the electron velocity distribution function
(EVDF) $F$
\begin{equation}
\frac{\partial F}{\partial t}+v_{x}\frac{\partial F}{\partial x}-\frac{e}{m}%
E_{x}\frac{\partial F}{\partial v_{x}}=0,  \label{Vlasov general}
\end{equation}%
together with the Poisson equation
\begin{equation}
\frac{dE}{dx}=4\pi e\left( n_{i}-\int_{-\infty }^{\infty }Fdv_{x}\right) .
\label{Poisson apend}
\end{equation}%
In the linear approximation, the EVDF can be split into two parts
\begin{equation}
F(t,x,v_{x})=f_{0}(v_{x})+f(t,x,v_{x}),  \label{linearization}
\end{equation}%
where $f_{0}(v_{x})$ describes EVDF of a uniform plasma with uniform ion
density $n_{e}=n_{i}=n_{0}$ and $f(t,x,v_{x})$ is EVDF due a wave
perturbation. Substituting Eq.~(\ref{linearization}) into Eqs.~(\ref{Vlasov
general}) and (\ref{Poisson apend}) yields the linearized Vlasov-Poisson
system of equations
\begin{equation}
\frac{\partial f}{\partial t}+v_{x}\frac{\partial f}{\partial x}-\frac{e}{m}%
E_{x}\frac{df_{0}}{dv_{x}}=-\nu f,  \label{Vlasov linear 1}
\end{equation}%
\begin{equation}
\frac{dE_{x}}{dx}=-4\pi e\int_{-\infty }^{\infty }f(v_{x})dv_{x}.
\label{Poisson linear 1}
\end{equation}%
In the first equation (\ref{Vlasov linear 1}), the small collisional term
with the collision frequency $\nu \ll \omega $ is taken into account. In
Ref.~\cite{Landau} Landau solved the linearized Vlasov-Poisson system making
use of the Laplace transform for a semi-infinite plasma $x>0$. However, it
is more convenient to apply a Fourier transform to an infinite plasma by
artificially continuing the EVDF and the electric field in the semi-plane $%
x<0$ \cite{Aliev and me}. Electrons moving with $v_{x}<0$ reflect from the
boundary $x=0$ and change their velocity to $-v_{x}$. This gives the
boundary condition for the Vlasov equation in the semi-plane $x>0$
\begin{equation}
f(t,0,v_{x})=f(t,0,-v_{x}).  \label{Vlasov bc}
\end{equation}%
Instead of considering problem in the semi-plane $x>0$ with the boundary
condition in Eq.(\ref{Vlasov bc}), we can consider the entire plane $x\in
\lbrack -\infty ,\infty ]$ by artificially continuing the electric field
into the semi-plane $x<0$. The Vlasov equation is symmetric with respect to
a change in variables according to the substitution
\begin{equation}
v_{x}\rightarrow -v_{x},\;x\rightarrow -x,\;E\rightarrow -E.
\end{equation}%
Therefore, electrons at $x=0$ with $v_{x}>0$, which are reflected from the
wall can be represented as electrons which came from the semi-plane $x<0$
and interacted with the electric field
\begin{equation}
E_{x}(x<0)=-E_{x}(x>0).  \label{E symmetry}
\end{equation}%
As a result, the electric field has to be continued anti-symmetrically into
the semi-plane $x<0$.

Now we can apply the Fourier transform for the Vlasov-Poisson system of
Eqs.~(\ref{Vlasov linear 1}) and (\ref{Poisson linear 1}). This gives for
the components of the EVDF $f_{k}e^{-i\omega t+kx}$ and the electric field $%
E_{k}e^{-i\omega t+kx}$
\begin{equation}
-i(\omega +i\nu -v_{x}k)f_{k}-\frac{e}{m}E_{k}\frac{df_{0}}{dv_{x}}=0,
\label{Vlasov linear k}
\end{equation}%
\begin{equation}
ikE_{k}+2E_{0}=4\pi e\int_{-\infty }^{\infty }f_{k}dv_{x}.
\label{Poisson linear k}
\end{equation}%
Note that due the fact that the electric field is a discontinuous function,
the Fourier transform of the derivative of the electric field $dE/dx$ is $%
ikE+2E_{0}$, where $E_{0}=E(0)$ is the electric field at the right side ($x>0
$) of the plasma boundary. Substituting $f_{k}$ from Eq.(\ref{Vlasov linear
k}) into (\ref{Poisson linear k}) yields
\begin{equation}
E_{k}=\frac{2E_{0}}{ik}\frac{1}{\varepsilon _{\Vert }(\omega ,k)},
\label{Ek}
\end{equation}%
where $\varepsilon _{\Vert }(\omega ,k)$ is the longitudinal plasma
permittivity
\begin{equation}
\varepsilon _{\Vert }(\omega ,k)=1+\frac{\omega _{p}^{2}}{n_{0}k}%
\int_{-\infty }^{\infty }\frac{1}{\omega +i\nu -v_{x}k}\frac{df_{0}}{dv_{x}}%
dv_{x}.  \label{dielectric function}
\end{equation}%
Substituting a Maxwellian EEDF
\begin{equation}
f_{0}=\frac{n_{0}}{\sqrt{\pi }v_{T}}\exp (-v^{2}/v_{T}^{2}),
\end{equation}%
where $v_{T}=\sqrt{2T/m}$, into Eq.(\ref{dielectric function}) and after
some algebra \cite{Lifshitz and Pitaevskii}, we obtain
\begin{equation}
\varepsilon _{\Vert }(\omega ,k)\simeq 1+\frac{2\omega _{p}^{2}}{%
k^{2}v_{T}^{2}}\left[ 1+\frac{1}{\sqrt{\pi }v_{T}}\int_{-\infty }^{\infty }%
\frac{\exp (-v^{2}/v_{T}^{2})}{v_{x}k-\omega -i\nu }dv_{x}\right] .
\label{dielectric function 1}
\end{equation}%
The last term on the right hand side can be expressed in terms of the plasma
dispersion function
\begin{equation}
Z(\zeta )=\frac{1}{\sqrt{\pi }}\int_{-\infty }^{\infty }\frac{\exp (-t^{2})}{%
t-\zeta }dt,\,~~\;Im(\zeta )>0.  \label{Dispersion function}
\end{equation}%
The dispersion function $Z(\zeta )$ in the form of Eq. (\ref{Dispersion
function}) is only defined for $Im(\zeta )>0$ and is defined as an
analytical continuation for $Im(\zeta )<0$. For $k>0$, in the limit $\nu
\rightarrow 0$,
\begin{equation}
\frac{1}{\sqrt{\pi }v_{T}}\int_{-\infty }^{\infty }\frac{\exp
(-v^{2}/v_{T}^{2})}{v_{x}k-\omega -i\nu }dv_{x}=\frac{1}{kV_{T}/\omega }%
Z(\omega /kV_{T}).
\end{equation}%
For $k<0$, the imaginary part of the $(\omega +i\nu )/k$ is negative and we
have to transform the integral (\ref{dielectric function 1}) so that the
pole $v_{xp}=(\omega +i\nu )/k$ lies in the upper plane of the complex
velocity. This can be achieved by substitution $-v_{x}\rightarrow v_{x}$ ,
which gives for $k<0$
\begin{equation}
\frac{1}{\sqrt{\pi }v_{T}}\int_{-\infty }^{\infty }\frac{\exp
(-v^{2}/v_{T}^{2})}{v_{x}|k|-\omega -i\nu }dv_{x}=\frac{1}{|k|v_{T}/\omega }%
Z(\omega /|k|v_{T}).
\end{equation}%
As a result,
\begin{equation}
\varepsilon _{\Vert }(\omega ,k)\simeq 1+\frac{2\omega _{p}^{2}}{%
k^{2}v_{T}^{2}}\left[ 1+\frac{1}{|kV_{T}/\omega |}Z(|\omega /kV_{T}|)\right]
.  \label{dielectric function 2}
\end{equation}%
Note that because the function $f_{0}$ is symmetric with respect to the
substitution $-v_{x}\rightarrow v_{x}$, $\varepsilon (\omega ,k)$ is
symmetric with respect to the substitution $-k\rightarrow k$.
Correspondingly the symmetry of the electric field in Eq.(\ref{E symmetry})
is preserved.

The electric field profile is given by the inverse Fourier transform of Eq.( %
\ref{Ek})
\begin{equation}
E_{x}(x)=\frac{1}{2\pi }\int_{-\infty }^{\infty }\frac{2E_{0}}{ik}\frac{
e^{ikx}}{\varepsilon _{\Vert }(\omega ,k)}dk.
\label{Efield inverse Fourier transform CCP}
\end{equation}

In the limit $x\rightarrow \infty $, $E(x)\rightarrow E_{0}/\varepsilon $,
where $\varepsilon =\varepsilon (\omega ,0)$. This is in accord with the
conservation of the total current in the one-dimensional geometry. The total
current is the sum of the displacement current and the electron current,
\begin{equation}
\frac{1}{4\pi }\frac{\partial E_{x}}{\partial t}+j_{e}=I(t).
\label{conservation of I}
\end{equation}%
The total current conservation follows from the combination of the Poisson
equation and the charge continuity equation. Indeed, taking the time
derivative of the Poisson equation and making use of the charge continuity
equation gives
\begin{equation}
\nabla \cdot \frac{\partial }{\partial t}E_{x}+4\pi \nabla \cdot j_{e}=0.
\end{equation}%
In one-dimensional geometry it can be integrated with a constant of space --
the total current carrying through the plasma $I(t)$, which gives Eq.(\ref%
{conservation of I}). For a harmonic electric field considered here, Eq.(\ref%
{conservation of I}) gives
\begin{equation}
-i\omega E_{x}-4\pi i\omega \left( \frac{\varepsilon -1}{4\pi }\right)
E_{x}=-i\omega E_{0}.  \label{conservation of I in eps}
\end{equation}%
Here, we account for the relationship between the plasma conductivity ($%
j_{e}=\sigma E$) and the plasma dielectric function $\varepsilon =1+4\pi
\sigma /(-i\omega )$. Eq.(\ref{conservation of I in eps}) gives
\begin{equation}
E_{x}(x\rightarrow \infty )=E_{0}/\varepsilon .
\end{equation}%
The same result can be obtained from Eq.(\ref{Efield inverse Fourier
transform CCP}) after substituting $\varepsilon (\omega ,k)\rightarrow
\varepsilon (\omega ,0)$ and integrating. Thus, the electric field in the
transition region is given by
\begin{equation}
E_{x}(x)-E_{0}/\varepsilon =\frac{1}{2\pi }\int_{-\infty }^{\infty }\frac{%
2E_{0}}{ik}\left( \frac{1}{\varepsilon _{\Vert }(\omega ,k)}-\frac{1}{%
\varepsilon _{\Vert }(\omega ,0)}\right) e^{ikx}dk.
\end{equation}%
The dielectric function in the form given by Eq.(\ref{dielectric function 1}%
) is not an analytic function of $k$. To apply the theory of residues,
Landau proposed to split integral into two parts \cite{Landau} according to
\begin{eqnarray}
E_{x}(x)-E_{0}/\varepsilon  &=&\frac{1}{2\pi }\int_{-\infty }^{\infty }\frac{%
2E_{0}}{ik}\left( \frac{1}{\varepsilon _{1}(\omega ,k)}-\frac{1}{\varepsilon
(\omega ,0)}\right) e^{ikx}dk \\
&&+\frac{1}{2\pi }\int_{0}^{\infty }\frac{2E_{0}}{ik}\left( \frac{1}{%
\varepsilon _{1}(\omega ,k)}-\frac{1}{\varepsilon _{\Vert }(\omega ,k)}%
\right) e^{ikx}dk,  \nonumber
\label{Efield
inverse Fourier transform landau}
\end{eqnarray}%
where
\begin{equation}
\varepsilon _{1}(\omega ,k)=1+\frac{2\omega _{p}^{2}}{\omega ^{2}k^{2}}\left[
1-\frac{1}{kv_{T}/\omega }Z(-\omega /kv_{T})\right] .
\end{equation}

The first integral can be calculated by moving the path of integration into
the complex $k-$plane and applying the theory of residues. For $\omega \ll
\omega _{p}$, $\varepsilon <0$ and there is only one pole $\varepsilon
_{1}(\omega ,k)=0$ in the upper half-plane \cite{Landau}. It corresponds to
the usual screening with the Debye length. In the limit $k\sim \omega
_{p}/v_{T},$ $Z(|\omega /kv_{T}|)\sim 1$ and $Z(|\omega
/kv_{T}|)/|kv_{T}/\omega |\ll 1$, which gives
\begin{equation}
\varepsilon _{1}(\omega ,k)\simeq 1+\frac{2\omega _{p}^{2}}{k^{2}v_{T}^{2}}.
\end{equation}%
Calculation of the first term in Eq.(\ref{Efield inverse Fourier transform
landau}) gives $E_{0}\exp (-x/a)$, where $a$ is the Debye length $a=v_{T}/%
\sqrt{2}\omega _{p}$. Therefore,
\begin{equation}
E_{x}(x)=E_{0}/\varepsilon +E_{0}\exp (-x/a)+\frac{1}{2\pi }\int_{0}^{\infty
}\frac{2E_{0}}{ik}\left( \frac{\varepsilon _{\Vert }(\omega ,k)-\varepsilon
_{1}(\omega ,k)}{\varepsilon _{1}(\omega ,k)\;\varepsilon _{\Vert }(\omega
,k)}\right) e^{ikx}dk  \label{Efield inverse Fourier transform 2}
\end{equation}%
For $Im(k)=0$, $Z(-\omega /kv_{T})=-Z(\omega /kv_{T})^{\ast }$ \cite{Plasma
Formulary} and
\begin{equation}
\varepsilon _{1}(\omega ,k)=1+\frac{2\omega _{p}^{2}}{v_{T}^{2}k^{2}}\left[
1+\frac{1}{kv_{T}/\omega }Z(\omega /kv_{T})^{\ast }\right] .
\label{dielectric function 3}
\end{equation}%
Substituting Eq.(\ref{dielectric function 3}) into Eq.(\ref{Efield inverse
Fourier transform 2}) gives for the last term $E_{t}(x)$%
\begin{equation}
E_{t}(x)=\frac{4E_{0}}{\pi }\frac{\omega \omega _{p}^{2}}{v_{T}^{3}}%
\int_{0}^{\infty }\frac{1}{k^{4}}\frac{Im[Z(\omega /kv_{T})]}{\varepsilon
_{1}(\omega ,k)\;\varepsilon _{\Vert }(\omega ,k)}e^{ikx}dk,  \label{Et}
\end{equation}%
where \cite{Plasma Formulary}
\begin{equation}
Im[Z(\zeta )]=\sqrt{\pi }\exp (-\zeta ^{2}).  \label{Im Z}
\end{equation}%
The last integral can be calculated analytically only in the limit $x\gg
v_{T}/\omega $ by applying the method of steepest descend. In this limit, $%
k\ll v_{T}/\omega ,$ $\varepsilon _{1}(\omega ,k)\approx \varepsilon (\omega
,k)\approx \varepsilon $ and
\begin{equation}
\int_{0}^{\infty }\frac{1}{k^{4}}\exp (ikx-\omega
^{2}/k^{2}v_{T}^{2})dk\simeq \frac{\sqrt{2\pi }}{\sqrt{3}}(x\lambda _{\omega
})^{2/3}\lambda _{\omega }\exp \left[ c\left( \frac{x}{\lambda _{\omega }}%
\right) ^{2/3}-i\pi /3\right] ,  \label{integral1}
\end{equation}%
where $c=3(-1+i\sqrt{3})/4$, and $\lambda _{\omega }=v_{T}/\sqrt{2}\omega $
is the phase-mixing scale.

Substituting Eq.(\ref{Im Z}) into Eq.(\ref{Et}) and making use of Eq.(\ref%
{integral1}) yields at $x\gg \lambda _{\omega }$ \cite{Landau}
\begin{equation}
E_{t}(x)\approx \frac{2E_{0}}{\sqrt{3}\varepsilon ^{2}}\frac{\omega _{p}^{2}
}{\omega ^{2}}\left( \frac{x}{\lambda _{\omega }}\right) ^{2/3}\exp \left[
c\left( \frac{x}{\lambda _{\omega }}\right) ^{2/3}-i\pi /3\right] .
\label{Et landau}
\end{equation}

The plots of amplitude and phase of the electric field profile $E_{t}(x)$
given by Eq.(\ref{Et}) and the approximate analytical result Eq.~(\ref{Et
landau}) are shown in Fig.~\ref{Figure Et}.

\section{Analytical derivation of the transverse rf electric field profile
near the plasma boundary ($\mathbf{E\perp k}$)}

The analytical solution involves solving the Vlasov equation for the
electron velocity distribution function (EVDF) $F$%
\begin{equation}
\frac{\partial F}{\partial t}+v_{x}\frac{\partial F}{\partial x}-\frac{e}{m}%
(E_{y}+v_{x}\times B_{z})\frac{\partial F}{\partial v_{y}}=0.
\label{Vlasov general Et}
\end{equation}%
This equation has to be solved together with the Maxwell's equation yielding
\begin{equation}
\left( \frac{d^{2}}{dx^{2}}+\frac{\omega ^{2}}{c^{2}}\right) E_{y}\mathbf{=-}%
\frac{4\pi i\omega }{c^{2}}\left[ j+I\delta (x)\right] ,
\label{Maxwell 1D append}
\end{equation}%
where $I$ is the surface current. The plasma density is not perturbed in the
transverse wave; therefore there is no need to solve the Poisson equation.
In the linear approximation, the EVDF can be split into two parts
\begin{equation}
F(t,x,\mathbf{v})=f_{0}(v)+f(t,x,\mathbf{v}),  \label{linearization 2}
\end{equation}%
where $f_{0}(v)$ describes EVDF of an isotropic, uniform plasma with uniform
ion density $n_{e}=n_{i}=n_{0}$ and $f(t,x,\mathbf{v})$ is the EVDF due a
wave perturbation. Substituting Eq.~(\ref{linearization 2}) into Eqs.~(\ref%
{Vlasov general Et}) yields the linearized Vlasov equation
\begin{equation}
\frac{\partial f}{\partial t}+v_{x}\frac{\partial f}{\partial x}-\frac{e}{m}%
E_{y}\frac{\partial f_{0}}{\partial v_{y}}=-\nu f.  \label{Vlasov linear Ey}
\end{equation}%
In Eq.~(\ref{Vlasov linear Ey}), the small collisional term with collision
frequency $\nu \ll \omega $ is taken into account. Similarly to the case of
the longitudinal wave, we can consider the entire plane $x\in \lbrack
-\infty ,\infty ]$ by artificially continuing the electric field in the
semi-plane $x<0$. The Vlasov equation is symmetric relative to the change in
variables according to the substitution
\begin{equation}
v_{x}\rightarrow -v_{x},\;x\rightarrow -x,\;E_{y}\rightarrow E_{y}.
\end{equation}%
Therefore, electrons at $x=0$ with $v_{x}>0$ which are reflected from the
wall can be represented as electrons which came from the semi-plane $x<0$
and interacted with the electric field
\begin{equation}
E_{y}(x<0)=E_{y}(x>0).  \label{E symmetry Et}
\end{equation}%
As a result, the electric field has to be continued symmetrically into the
semi-plane $x<0$.

Now we can apply the Fourier transform for Eqs.~(\ref{Vlasov linear Ey}) and
(\ref{Maxwell 1D append}). This gives for components of the EVDF $%
f_{k}e^{-i\omega t+ikx}$ and the electric field $E_{yk}e^{-i\omega t+ikx}$
\begin{equation}
-i(\omega +i\nu -v_{x}k)f_{k}-\frac{e}{m}E_{yk}\frac{\partial f_{0}}{%
\partial v_{y}}=0,  \label{Vlasov linear k Et}
\end{equation}%
\begin{equation}
\left( -k^{2}+\frac{\omega ^{2}}{c^{2}}\right) E_{yk}\mathbf{=-}\frac{4\pi
i\omega }{c^{2}}(j_{k}+I).  \label{Maxwell Etk}
\end{equation}%
Substituting $f_{k}$ from Eq.~(\ref{Vlasov linear k Et}) into (\ref{Maxwell
Etk}) with the current $j_{k}=-e\int f_{k}v_{y}d\mathbf{v}$ yields
\begin{equation}
E_{yk}=\frac{4\pi i\omega }{c^{2}}I\frac{1}{k^{2}-\frac{\omega ^{2}}{c^{2}}%
\varepsilon _{t}(\omega ,k)},  \label{Etk}
\end{equation}%
where $\varepsilon _{t}(\omega ,k)$ is the transverse plasma permittivity
\begin{equation}
\varepsilon _{t}(\omega ,k)=1+\frac{\omega _{p}^{2}}{n_{0}\omega }%
\int_{-\infty }^{\infty }\frac{v_{y}}{\omega +i\nu -v_{x}k}\frac{\partial
f_{0}}{\partial v_{y}}dv_{x}.  \label{dielectric function Et}
\end{equation}%
Substituting a Maxwellian EEDF gives \cite{Lifshitz and Pitaevskii}
\begin{equation}
\varepsilon _{t}(\omega ,k)=1+\frac{\omega _{p}^{2}}{\omega ^{2}}\frac{%
\omega }{v_{T}|k|}Z\left( \frac{\omega }{v_{T}|k|}\right) .
\label{dielectric function 2 Et}
\end{equation}%
Note that because the function $f_{0}$ is symmetric relative to the
substitution $-v_{x}\rightarrow v_{x}$, $\varepsilon (\omega ,k)$ is
symmetric relative to the substitution $-k\rightarrow k$. Correspondingly,
the symmetry of the electric field in Eq.~(\ref{E symmetry Et}) is preserved.

The electric field profile is given by the inverse Fourier transform of Eq.(%
\ref{Etk})
\begin{equation}
E_{y}(x)=\frac{1}{2\pi }\int_{-\infty }^{\infty }\frac{4\pi i\omega }{c^{2}}I%
\frac{e^{ikx}}{k^{2}-\omega ^{2}\varepsilon _{t}(\omega ,k)/c^{2}}dk.
\label{Efield inverse Fourier transform}
\end{equation}%
Similar to the analysis of the longitudinal wave, we split the integral in
Eq.~(\ref{Etk}) into two parts
\begin{equation}
E_{y}(x)=E_{yp}(x)+E_{yt}(x),
\end{equation}%
where
\begin{equation}
E_{yp}(x)=\frac{2\pi i\omega }{c^{2}}I\int_{-\infty }^{\infty }\frac{e^{ikx}%
}{k^{2}-\omega ^{2}\varepsilon _{t1}(\omega ,k)/c^{2}}dk,
\end{equation}%
and
\begin{equation}
E_{yt}(x)=\frac{2i\omega }{c^{2}}I\frac{\omega ^{2}}{c^{2}}\int_{0}^{\infty }%
\frac{\left[ \varepsilon _{t}(\omega ,k)-\varepsilon _{t1}(\omega ,k)\right]
e^{ikx}}{\left[ k^{2}-\omega ^{2}\varepsilon _{t1}(\omega ,k)/c^{2}\right] %
\left[ k^{2}-\omega ^{2}\varepsilon _{t}(\omega ,k)/c^{2}\right] }dk,
\end{equation}%
\begin{equation}
\varepsilon _{t1}(\omega ,k)=1-\frac{\omega _{p}^{2}}{\omega ^{2}}\frac{%
\omega }{v_{T}k}Z\left( -\frac{\omega }{v_{T}k}\right) .
\end{equation}%
Note that $\varepsilon _{t1}(\omega ,k)=\varepsilon _{t}(\omega ,k)$ for $k<0
$.

The first part $E_{yp}(x)$ of the electric field can be calculated by
evaluating the integral in the complex $k-$plane. A pole of $E_{yp}(x)$- $%
k_{p}i$ lies on the imaginary axis of the $k-$plane. The dielectric
permittivity is real and negative on imaginary axis of the $k-$plane
\begin{equation}
\varepsilon _{t1}(\omega ,k_{p}i)=1-\frac{\omega _{p}^{2}}{\omega ^{2}}\frac{%
\omega }{v_{T}k_{p}}F\left( \frac{\omega }{v_{T}k_{p}}\right) ,
\end{equation}%
where $F(\zeta )=ImZ\left( i\zeta \right) =\sqrt{\pi }\exp (y^{2})erfc(y)$
\cite{Plasma Formulary}. There is always a real value of $k_{p}$ as the root
of
\begin{equation}
k_{p}^{2}=-\omega ^{2}\varepsilon _{t1}(\omega ,ik_{p})/c^{2}.
\end{equation}%
Applying the theory of residues, the integral for $E_{yp}(x)$ gives
\begin{equation}
E_{yp}(x)=E_{op}e^{-k_{p}x},
\end{equation}%
where
\begin{equation}
E_{op}=\frac{2i\omega }{c^{2}}\frac{2\pi iI}{2k_{p}i-d\varepsilon
_{t1}(\omega ,k)/dk\;\omega ^{2}/c^{2}}.
\end{equation}

In the limit $x\gg \delta $, the last term $E_{yt}(x)$ can be calculated
making use of the method of steepest descend. Substituting $k^{2}-\omega
^{2}\varepsilon _{t1}(\omega ,k)/c^{2}$ in the denominator of the expression
for $E_{yt}(x)$ by its limit $=\omega _{p}^{2}/c^{2}$ at $k\rightarrow 0$,
gives
\begin{equation}
E_{yt}(x)=-\frac{4\omega ^{2}\sqrt{\pi }}{\omega _{p}^{2}v_{T}}I\left[
\int_{0}^{\infty }\frac{1}{k}\exp \left[ -\left( \frac{\omega }{v_{T}k}%
\right) ^{2}+ikx\right] dk\right] ,
\end{equation}%
which yields
\begin{equation}
E_{ystd}(x)=-\frac{4\omega ^{2}\pi }{\omega _{p}^{2}v_{T}}I\frac{\sqrt{2}}{%
\sqrt{3}}\left( \frac{x}{\lambda _{\omega }}\right) ^{-1/3}\exp \left[
c\left( \frac{x}{\lambda _{\omega }}\right) ^{2/3}-i\pi /2\right] ,
\label{std anomolous skin}
\end{equation}%
where $c=3(-1+i\sqrt{3})/4$ and $\lambda _{\omega }=v_{T}/\sqrt{2}\omega $.


\newpage

\end{document}